\documentclass[12pt,english]{aastex}
\usepackage{mathptmx}

\usepackage{courier}

\usepackage[T1]{fontenc}
\usepackage[latin9]{inputenc}
\usepackage{babel}
\usepackage{amssymb}
\usepackage{esint}
\usepackage[unicode=true,pdfusetitle,
 bookmarks=true,bookmarksnumbered=false,bookmarksopen=false,
 breaklinks=false,pdfborder={0 0 1},backref=false,colorlinks=false]
 {hyperref}
\usepackage{breakurl}

\makeatletter

\usepackage{hyperref}

\makeatother

\begin{document}

\title{Cosmic Ray Transport with Magnetic Focusing and the ``Telegraph''
model}

\author{M.A. Malkov$^{1}$, R.Z. Sagdeev$^{2}$}

\affil{$^{1}$CASS and Department of Physics, University of California,
San Diego, La Jolla, CA 92093\\$^{2}$University of Maryland, College
Park, MD 20742-3280}
\begin{abstract}
Cosmic rays (CR), constrained by scattering on magnetic irregularities,
are believed to propagate diffusively. But a well-known defect of
diffusive approximation whereby some of the particles propagate unrealistically
fast has directed interest towards an alternative CR transport model
based on the ``telegraph'' equation. However, its derivations often
lack rigor and transparency leading to inconsistent results.

We apply the classic Chapman-Enskog method to the CR transport problem.
We show that no ``telegraph'' (second order time derivative) term
emerges in any order of a proper asymptotic expansion with systematically
eliminated short time scales. Nevertheless, this term may formally
be\emph{ converted} from the \emph{fourth} order hyper-diffusive term
of the expansion. But, both the telegraph and hyperdiffusive terms
may only be important for a short relaxation period associated with
either strong pitch-angle anisotropy or spatial inhomogeneity of the
initial CR distribution. Beyond this period the system evolves diffusively
in both cases. The term conversion, that makes the telegraph and Chapman-Enskog
approaches reasonably equivalent, is possible only after this relaxation
period. During this period, the telegraph solution is argued to be
unphysical. Unlike the hyperdiffusion correction, it is not uniformly
valid and introduces implausible singular components to the solution.
These dominate the solution during the relaxation period. As they
are shown not to be inherent in the underlying scattering problem,
we argue that the telegraph term is involuntarily acquired in an asymptotic
reduction of the problem.
\end{abstract}

\section{Preliminary Considerations\label{sec:Preliminary-Considerations}}

The problem addressed here is fundamental and not new to the cosmic
ray (CR) transport studies. It can be formulated very plainly: How
to describe CR transport by only their isotropic component, after
the anisotropic one has decayed by scattering on magnetic irregularities?
Suppose the angular distribution of CRs is given by the function $f\left(\mu,t,z\right)$
obeying an equation from which the rapid gyro-phase rotation is already
removed (drift approximation, e.g., \citealp{VVSQL62,Kulsrud05})

\begin{equation}
\frac{\partial f}{\partial t}+v\mu\frac{\partial f}{\partial z}=\frac{\partial}{\partial\mu}\left(1-\mu^{2}\right)\mathcal{D}\left(\mu\right)\frac{\partial f}{\partial\mu}.\label{eq:PAscatIntro}
\end{equation}
Here $z$ is the local coordinate along the ambient magnetic field,
$\mu$ is the cosine of the particle pitch angle, and $\mathcal{D}$
is the pitch angle diffusion coefficient. Now, we make the next step
in simplifying the transport description and seek an equation for
the pitch-angle averaged distribution

\[
f_{0}\left(t,z\right)\equiv\frac{1}{2}\intop_{-1}^{1}f\left(\mu,t,z\right)d\mu\equiv\left\langle f\right\rangle .
\]
The basic solution to this problem has been known for at least half
a century (e.g., \citealp{Jokipii66} and references therein). To
the leading order in $1/\mathcal{D}$ (assuming the characteristic
scale and time of the problem being longer than particle mean free
path and collision time) it can be obtained straightforwardly by averaging
eq.(\ref{eq:PAscatIntro})

\[
\frac{\partial f_{0}}{\partial t}=-\frac{v}{2}\frac{\partial}{\partial z}\left\langle \left(1-\mu^{2}\right)\frac{\partial f}{\partial\mu}\right\rangle ,
\]
and substituting $\partial f/\partial\mu\ll f_{0}$ from eq.(\ref{eq:PAscatIntro})
as:

\begin{equation}
\frac{\partial f}{\partial\mu}\approx-\frac{v}{2\mathcal{D}}\frac{\partial f_{0}}{\partial z}.\label{eq:dfdmuIntro}
\end{equation}
Thus, the following diffusion equation results for $f_{0}$:

\begin{equation}
\frac{\partial f_{0}}{\partial t}=\frac{v^{2}}{4}\frac{\partial}{\partial z}\left\langle \frac{1-\mu^{2}}{\mathcal{D}}\right\rangle \frac{\partial f_{0}}{\partial z}.\label{eq:DifIntro}
\end{equation}

A questionable point of course is neglecting $\partial f/\partial t$
in favor for $v\partial f/\partial z$ in eq.(\ref{eq:dfdmuIntro}).
It is somewhat justified by the small parameter $\mathcal{D}^{-1}\ll1$
in the \emph{final result}, given by eq.(\ref{eq:DifIntro}), making
$\partial f/\partial t$ hopefully small. On the other hand this is
true for $\partial f_{0}/\partial t$ but not necessarily for $\partial f/\partial t$,
since the latter may contain also the rapidly decaying anisotropic
part $\tilde{f}=f-f_{0}$ of the initial CR distribution. For $\mathcal{D}t\gtrsim1$,
however, $\tilde{f}$ must die out and neglecting $\partial f/\partial t$
appears plausible for the \emph{long-term} CR transport. At the same
time, $\partial f_{0}/\partial t$ is large when $f_{0}$ is very
narrow in $z$ initially, such as in the fundamental solution. In
the sequel, these aspects of the CR propagation will be a key for
choosing an appropriate asymptotic reduction method.

However convincing the justification, the CR diffusion model encounters
the problem of a superluminal, or simply ``too-fast'' particle propagation.
Although rather common for diffusive models, the problem is largely
ignorable as long as the number of such particles remains small. There
are cases, however, such as the propagation of ultra high-energy cosmic
rays, where this problem must be addressed \citep{AloisBerezSuperLum09}.
Various attempts, starting as early as in 60s, e.g., \citep{AxfordTelegr65},
have been made to devise a better transport equation for CRs. Unfortunately,
in our view, they lack mathematical rigor and clarity and sometimes
lead to inconsistent results. 

In the most recent telegraph model, due to \citet{LitvSchlick13},
a higher order in $1/\mathcal{D}\ll1$ term was included by retaining
$\partial f/\partial t$, dropped in the simplest derivation above.
This strategy gave rise to an additional $\partial^{2}f_{0}/\partial t^{2}$
-term in the ``master'' equation. This additional term transforms
eq.(\ref{eq:DifIntro}) into a ``telegraph'' equation: 

\begin{equation}
\frac{\partial f_{0}}{\partial t}+T\frac{\partial^{2}f_{0}}{\partial t^{2}}=\frac{\partial}{\partial z}k\frac{\partial f_{0}}{\partial z}+\frac{k}{L}\frac{\partial f_{0}}{\partial z}\label{eq:TelegrEqIntro}
\end{equation}
with

\begin{equation}
k=\frac{v^{2}}{4}\left\langle \frac{1-\mu^{2}}{\mathcal{D}}\right\rangle ,\;\;\; T=\left.\left\langle \left[\intop_{0}^{\mu}d\mu/\mathcal{D}\right]^{2}\right\rangle \middle/\left\langle \frac{1-\mu^{2}}{\mathcal{D}}\right\rangle \right.,\;\;\; L^{-1}=-B^{-1}\partial B/\partial z\label{eq:KappaTau}
\end{equation}
For the sake of comparison with earlier telegraph equation results
that will be made in Sec.\ref{sec:Comparison}, we have added here
the magnetic focusing effect (the last term on the r.h.s. with $B$$\left(z\right)$
being the magnetic field), not included initially in eq.(\ref{eq:PAscatIntro}).
Eq.(\ref{eq:TelegrEqIntro}) is just a linear equation that can be
solved immediately. The fundamental solution to eq.(\ref{eq:TelegrEqIntro})
that starts off from a $\delta\left(z\right)$ distribution, instantaneously
released at $t=0$, is as follows (e.g., \citealp{GOLDSTEIN01011951,AxfordTelegr65,SchwadronTelegraph94,LitvEffenSchlick15},
$L^{-1}=0$, for simplicity)

\begin{eqnarray}
f_{0} & = & \frac{1}{2}e^{-t/2T}\left\{ \delta\left(z-\sqrt{\frac{k}{T}}t\right)+\delta\left(z+\sqrt{\frac{k}{T}}t\right)\right.\nonumber \\
 & + & \left.\frac{H\left(\sqrt{k/T}t-\left|z\right|\right)}{2\sqrt{kT}}\left[I_{0}\left(\frac{1}{2}\sqrt{\frac{t^{2}}{T^{2}}-\frac{z^{2}}{kT}}\right)+\frac{t}{\sqrt{T\left(kt^{2}-Tz^{2}\right)}}I_{1}\left(\frac{1}{2}\sqrt{\frac{t^{2}}{T^{2}}-\frac{z^{2}}{kT}}\right)\right]\right\} \label{eq:TelIntroSol}
\end{eqnarray}
Here $I_{0,1}$ denote the modified Bessel functions and $H$ - the
Heaviside unit function. 

One promising aspect of the telegraph equation is that it allows for
a ballistic mode of CR propagation when the initial conditions empower
the higher-order derivative terms to dominate (at least in early phase
of evolution). If, in addition, $T$ has a proper value, the bulk
speed of CRs may also be realistic. For example, this speed was derived
in \citep{Earl73} to be $v/\sqrt{3}$, which has also been used earlier
by \citealp{AxfordTelegr65}. This is just the rms velocity projection
of an isotropic, one-sided CR distribution on $z$-axis, which appears
to resolve the issue with the superluminal propagation. What is worrisome
here is that this bulk speed essentially requires a \emph{one-sided
}``isotropic'' CR distribution which, of course, is highly anisotropic
overall, contrary to the basic assumption of most treatments. So,
we need to start with isotropic initial distribution but, taking it
narrow and symmetric in $z$ (say Gaussian), the higher-order derivative
terms will, again, dominate in eq.(\ref{eq:TelegrEqIntro}) and the
single CR pulse will split into two, propagating in opposite directions
at the speeds $\pm\sqrt{k/T}$ (as $\partial f_{0}/\partial t=0$
at $t=0$, according to eq.{[}\ref{eq:PAscatIntro}{]}). Neglecting
$\partial f_{0}/\partial t$ (which is justified for $f_{0}$ sufficiently
narrow in $z$), the solution is simply $f_{0}\left(z,t\right)=F\left(z-\sqrt{k/T}t\right)+F\left(z+\sqrt{k/T}t\right)$,
where $2F\left(z\right)=f$$\left(z,0\right)$. This result casts
doubts on whether the telegraph term can be dominant under the assumption
of frequent CR scattering (asymptotic expansion in small $1/\mathcal{D}$),
as the initially sharp profile does not spread. Bringing $\partial f_{0}/\partial t$
back into the equation will only damp but not spread the profile,
as clearly seen from the solution given in eq.(\ref{eq:TelIntroSol}),
where $F\left(z\pm\sqrt{k/T}t\right)=\delta\left(z\pm\sqrt{k/T}t\right)$.
Besides, the bulk CR speed $\sqrt{k/T}=v/\sqrt{3}$ for an isotropic
scattering $\mathcal{D=}1$, although implicitly confirmed in the
recent derivation of the telegraph equation for an arbitrary $\mathcal{D}\left(\mu\right)$
by \citet{LitvSchlick13}, is not universally accepted. \citet{Gombosi93,Pauls93}
and \citet{SchwadronTelegraph94}, using simplified forms of $D\left(\mu\right)$,
advocate the value $\sqrt{5/11}v$ for the propagation speed. 

The last result is consistent with our calculations below, but with
strong reservations regarding the telegraph equation set out later
in the paper. Here we merely note that the solution of telegraph equation
specifically considered by \citet{LitvSchlick13}, which does not
have the property of splitting the initially narrow pulse into two,
does not conserve the total CR number $N=\intop f_{0}dz$. It starts
off from $N=0$ which is unphysical, as the equation has no source
on its r.h.s. To conserve $N$, two $\delta-$ pulses in eq.(\ref{eq:TelIntroSol})
are necessary. Those have been added to the treatment by \citet{LitvNoble13},
but the $\delta-$ pulses have not been shown on their plots, for
obvious reason. So, the comparison of this solution with the solution
of the original scattering problem is rather misleading. The disagreement
on the propagation speed $\sqrt{k/T}$ is also critical as the solution
in eq.(\ref{eq:TelIntroSol}) is cut off at a point $z$ moving with
this speed. For $z<\sqrt{k/T}t$ the profile is close to a Gaussian
(for $t\gg T$, where $T$ is the scattering time), so small variations
in the speed can produce significant variations in the solution. We
will also return to this later.

Another disadvantage of the telegraph equation (\ref{eq:TelegrEqIntro})
is that it is no longer an evolution equation and requires the time
derivative $\partial_{t}f_{0}$ as an initial condition. Although
this can be inferred from the angular distribution at $t=0$ using
eq.(\ref{eq:PAscatIntro}), the ``telegraph'' description of CR
transport is not self-contained. We show below that the $T$-term
in eq.(\ref{eq:TelegrEqIntro}) is subdominant in an asymptotic series
for $\mathcal{D}t\gtrsim1$, thus representing transients in the CR
transport. Strictly speaking, it should be omitted in the asymptotic
transport description along with the small hyper-diffusion term $\sim\partial^{4}f_{0}/\partial z^{4}$,
particularly if the term $\sim\partial^{3}f_{0}/\partial z^{3}$ does
not vanish. The latter was not included in eq.(\ref{eq:TelegrEqIntro}),
as it was obtained by applying insufficient direct iteration to eq.(\ref{eq:PAscatIntro})
in \citep{LitvSchlick13}, or assuming symmetric scattering, $\mathcal{D}\left(-\mu\right)=\mathcal{D}\left(\mu\right)$,
when it indeed vanishes. This symmetry restriction was relaxed in
\citep{Pauls93}. 

The reasons why we undertake the derivation of master equation to
a higher (fourth) order of approximation for an arbitrary $\mathcal{D}\left(\mu\right)$
are several. First, it is necessary to clarify the role of the telegraph
term entertained in the literature as an allegedly viable alternative
to the standard diffusion model. Second, it is important to obtain
the transport coefficients valid for arbitrary $\mathcal{D}\left(\mu\right)$,
that is for an arbitrary spectrum of magnetic fluctuations. As we
will show, the previous such derivation due to \citealp{LitvSchlick13},
does not include the third order term, while including only one fourth
order term, while there are more such terms (see eq.{[}\ref{eq:Master4}{]}).
Furthermore, the diffusion equation (\ref{eq:DifIntro}) supplemented
by a convective term $u\left(z\right)\partial f_{0}/\partial z$ for
the case of the bulk fluid (scattering center) motion with velocity
$u$, has long been and remains the main tool for diffusive shock
acceleration (DSA) models. An accurate assessment of the next non-vanishing
term, not included in eq.(\ref{eq:DifIntro}), is thus utterly important
for the DSA, particularly as claims are being made about the necessity
to include the telegraph term in the CR transport. In most DSA applications,
it is crucial to allow not only for an arbitrary fluctuation spectrum
$\mathcal{D}$ but for its dependence upon $f_{0}$ as well. This
dependence directly affects the particle spectrum and acceleration
time. We will discuss these aspects briefly in Sec.\ref{sec:Summary-and-Conclusions}.

In the next section, the basic transport equation with magnetic focusing
is introduced and the shortcomings of a reduction scheme based on
direct iterations are demonstrated. The appropriate asymptotic method
is elaborated in Sec.\ref{sec:Chapman-Enskog-expansion}. Apart from
what we already discussed regarding the telegraph equation, the objective
of Sec.\ref{sec:Chapman-Enskog-expansion} is to create a framework
suitable also for nonlinear (e.g., \citealp{PtuskinNLDIFF08,MDS10PPCF})
and quasi-linear \citep{Fujita11,MetalEsc13} versions of CR transport
which are important for both the DSA and for the subsequent escape
of the accelerated CR. In these settings, the CR pressure is high
enough to strongly modify at least the pitch-angle diffusion coefficient
$\mathcal{D}$ and possibly the shock structure itself \citep{MDS10PPCF}.
In Secs.\ref{sec:Comparison} and \ref{sec:Relation-between-Telegraph}
the implications of our results for the telegraph model and for the
long-time CR propagation are discussed, while Sec.\ref{sec:Summary-and-Conclusions}
concludes the paper.

\section{CR Transport Equation and its Asymptotic Reduction\label{sec:CR-Transport-Equation}}

Energetic particles (e.g., CRs) in a magnetic field, slowly varying
on the particle gyro-scale, are transported according to the following
gyro-phase averaged equation, e.g. \citep{VVSQL62,Jokipii66,Kulsrud05}

\begin{equation}
\frac{\partial f}{\partial t}+v\mu\frac{\partial f}{\partial z}+v\frac{\sigma}{2}\left(1-\mu^{2}\right)\frac{\partial f}{\partial\mu}=\frac{\partial}{\partial\mu}\nu D\left(\mu\right)\left(1-\mu^{2}\right)\frac{\partial f}{\partial\mu}\label{eq:Boltzmann}
\end{equation}
Here $v$ and $\mu$ are the particle velocity and pitch angle, $z$
points in the local field direction, $\sigma=-B^{-1}\partial B/\partial z$
is the magnetic field inverse scale and $\nu$ is the pitch angle
scattering rate, while $D\left(\mu\right)\sim1$ depends on the spectrum
of magnetic fluctuations. As the fastest transport is assumed to be
in $\mu$, we introduce the following small parameter

\begin{equation}
\varepsilon\equiv\frac{v}{l\nu}\equiv\frac{\lambda}{l}\ll1,\label{eq:EpsilonDef}
\end{equation}
where $\lambda$ is the particle mean free path and $l$ is a characteristic
scale that should be chosen depending on the problem considered. One
option is the scale of $B\left(z\right)$, in which case $l\sim\sigma^{-1}$.
If the CR source is present on the r.h.s. of eq.{[}\ref{eq:Boltzmann}{]}),
its scale can be taken as $l$. Finally, $l$ can be the scale of
an initial CR distribution. Strictly speaking, the shortest of these
scales should be taken as $l$. The problem with the initial CR distribution
is that in the most interesting case of the fundamental solution this
scale is zero. Therefore, over the initial period of CR spreading,
before the actual CR scale $l\left(t\right)\sim f/\left(\partial f/\partial z\right)$
exceeds the m.f.p. $\lambda$, direct asymptotic expansions in small
$\varepsilon$ remain inaccurate. The goal here is to choose the \emph{least
inaccurate} out of all possible expansion schemes. At a minimum, it
should be the one that does not introduce additional singularities,
apart from the initial delta function $\delta\left(z\right)$, that
must spread out under the particle recession and collisions. Therefore,
while taking $l=const$ in eq.(\ref{eq:EpsilonDef}), and assuming
$l\gg\lambda$, caution will be exercised during the initial phase
of the CR relaxation when the terms with higher spatial and time derivatives
are large, even if they contain small factors $\varepsilon^{n}\ll1$.
By measuring time in $\nu^{-1}$, $z$ in $l$, and simply replacing
$\sigma l\to\sigma\sim1$, the above equation transforms as follows

\begin{equation}
\frac{\partial f}{\partial t}-\frac{\partial}{\partial\mu}D\left(\mu\right)\left(1-\mu^{2}\right)\frac{\partial f}{\partial\mu}=-\varepsilon\left(\mu\frac{\partial f}{\partial z}+\frac{\sigma}{2}\left(1-\mu^{2}\right)\frac{\partial f}{\partial\mu}\right)\label{eq:BoltzmannRescaled}
\end{equation}

A suitable scheme for asymptotic reduction of the above equation using
$\varepsilon\ll1$ is due to Chapman and Enskog, suggested in development
of the earlier ideas by Hilbert (a good discussion of the history
of this method with mathematical details is given by \citealp{cercignani1988boltzmannequation}).
Originally, it was applied to Boltzmann equation in a strongly collisional
regime. Similar approaches have been used in plasma physics, e.g.,
in regards to the hydrodynamic description of collisional magnetized
plasmas \citep{Braginsky65} and the problem of run-away electrons
\citep{Gurevich61RunAway,KruskalBernstein64}.

Regardless of the asymptotic scheme, eq.(\ref{eq:BoltzmannRescaled})
suggests to seek $f$ as a series in $\varepsilon$

\begin{equation}
f=f_{0}+\varepsilon f_{1}+\varepsilon^{2}f_{2}+\dots\equiv f_{0}+\tilde{f}\label{eq:fAsSer}
\end{equation}
where

\begin{equation}
\left\langle f\right\rangle =f_{0},\ \ \ {\rm with}\ \ \ \left\langle \cdot\right\rangle =\frac{1}{2}\intop_{-1}^{1}\left(\cdot\right)d\mu,\label{eq:fAver}
\end{equation}
so that $\tilde{\left\langle f\right\rangle }=\left\langle f_{n>0}\right\rangle =0$.
The equation for $f_{0}$, which is the main (``master'') equation
of the method, takes the following form

\begin{equation}
\frac{\partial f_{0}}{\partial t}=-\varepsilon\left(\frac{\partial}{\partial z}+\sigma\right)\left\langle \mu f\right\rangle =-\frac{\varepsilon^{2}}{2}\left(\frac{\partial}{\partial z}+\sigma\right)\sum_{n=1}^{\infty}\varepsilon^{n-1}\left\langle \left(1-\mu^{2}\right)\frac{\partial f_{n}}{\partial\mu}\right\rangle \label{eq:Master}
\end{equation}
We see from this equation that, similarly to the case of Lorentz's
gas in an electric field \citep{Gurevich61RunAway,KruskalBernstein64},
$f_{0}$ depends on the ``slow time'' $t_{2}=\varepsilon^{2}t$
rather than on $t$. Indeed, the two problems are similar in that
they describe diffusive expansion of particles in phase space. The
expansion occurs in $z$-direction for the CR diffusion problem and
in energy for runaway electrons. The expansion is driven by a rapid
isotropization in pitch angle plus the convection in $z$- direction,
or acceleration in the electric field direction, for the CR transport
and electron runaway, respectively. 

The slow dependence of $f_{0}$ on time in eq.(\ref{eq:Master}) may
suggest to attribute the time derivative term in eq.(\ref{eq:BoltzmannRescaled})
to a higher order approximation (thus moving it to the r.h.s.). Such
ordering has been employed by \citet{LitvSchlick13} and the term
$\propto\partial^{2}f_{0}/\partial t^{2}$ has been produced in eq.(\ref{eq:Master}).
Obviously, a continuation of this process would result in progressively
higher time derivatives of $f_{0}$, corresponding to shorter and
shorter times in the initial relaxation. These transient phenomena
will be removed using the Chapman-Enskog asymptotic reduction scheme
in the next section. 

Unlike $f_{0}$, $\tilde{f}$ in eq.(\ref{eq:fAsSer}) does depend
on $t$ as on a ``fast'' time. Therefore, it is illegitimate to
attribute the first term on the l.h.s of eq.(\ref{eq:BoltzmannRescaled})
to any order of approximation different from that of the second term,
notwithstanding its fast decay for $t\gtrsim1$. Thus, using eqs.(\ref{eq:BoltzmannRescaled}-\ref{eq:fAsSer})
we must apply the following ordering

\begin{equation}
\frac{\partial f_{n}}{\partial t}-\frac{\partial}{\partial\mu}D\left(\mu\right)\left(1-\mu^{2}\right)\frac{\partial f_{n}}{\partial\mu}=-\mu\frac{\partial f_{n-1}}{\partial z}-\frac{\sigma}{2}\left(1-\mu^{2}\right)\frac{\partial f_{n-1}}{\partial\mu}\label{eq:InitOrdering}
\end{equation}
The above expansion scheme is sufficient to recover the leading order
of $f_{0}$ evolution from eq.(\ref{eq:Master}) by substituting there
$\partial f_{1}/\partial\mu\approx-\left(2D\right)^{-1}\partial f_{0}/\partial z$,
obtained from the last equation for $t\gtrsim1$. However, this scheme
is not suitable for determining $f_{n}$ for $n\ge2$ to submit to
eq.(\ref{eq:Master}). Indeed, as it may be seen from eq.(\ref{eq:InitOrdering}),
the solubility condition for $f_{2}$ at $t\gg1$ is $\left(\partial/\partial z+\sigma\right)\left\langle \left(1-\mu^{2}\right)\partial f_{1}/\partial\mu\right\rangle \approx-\left(\partial/\partial z+\sigma\right)\left\langle \left(1-\mu^{2}\right)/2D\right\rangle \partial f_{0}/\partial z=0$.
This is clearly too strong a restriction. The reason for this inconsistency
of the direct asymptotic expansion is that $f_{0}$ depends on time
much slower than $f_{n>0},$ so a slow time $t_{2}=\varepsilon^{2}t$
needs to be taken into consideration. The Chapman-Enskog method has
been developed for such cases, and we will make use of it in the next
section.

\section{Chapman-Enskog Expansion\label{sec:Chapman-Enskog-expansion}}

As we have seen, the asymptotic reduction of the original CR propagation
problem, given by eq.(\ref{eq:BoltzmannRescaled}), to its isotropic
part cannot proceed to higher orders of approximation using a simple
asymptotic series in eq.(\ref{eq:fAsSer}) and requires a multi-time
asymptotic expansion. In Chapman-Enskog method the operator $\partial/\partial t$
is expanded instead. Its purpose is to avoid unwanted higher time
derivatives to appear in higher orders of approximation. This is very
similar to, e.g., a secular growth in perturbed oscillations of dynamical
systems. To eliminate the secular terms, one seeks to alter (also
expand in small parameter) the frequency of the zero order motion,
which is similar to the $\partial/\partial t$ expansion. One example
of such approach may be found in a derivation of hydrodynamic equations
for strongly collisional but magnetized plasmas, starting from Boltzmann
equation \citep{MikhailovskyChEnsk67}. The classical monograph by
\citet{ChapmanCowling1991} (Ch.VIII) gives another example of a subdivision
of $\partial/\partial t$ operator for solving the transport problem
in a non-uniform gas-mixture. Expanding $\partial/\partial t$ operators
eliminates secular terms, such as the telegraph term. Perhaps more
customary today and equivalently is to introduce a hierarchy of formally
independent time variables (e.g., \citealp{Nayfeh81}) $t\to t_{0},\; t_{1},\dots$,
so that

\begin{equation}
\frac{\partial}{\partial t}=\frac{\partial}{\partial t_{0}}+\varepsilon\frac{\partial}{\partial t_{1}}+\varepsilon^{2}\frac{\partial}{\partial t_{2}}\dots\label{eq:TimeDerExpan}
\end{equation}
Instead of eq.(\ref{eq:InitOrdering}), from eq.(\ref{eq:BoltzmannRescaled})
we have

\begin{eqnarray}
\frac{\partial f_{n}}{\partial t_{0}}-\frac{\partial}{\partial\mu}D\left(\mu\right)\left(1-\mu^{2}\right)\frac{\partial f_{n}}{\partial\mu} & = & -\mu\frac{\partial f_{n-1}}{\partial z}-\frac{\sigma}{2}\left(1-\mu^{2}\right)\frac{\partial f_{n-1}}{\partial\mu}-\sum_{k=1}^{n}\frac{\partial f_{n-k}}{\partial t_{k}}\label{eq:fnGen}\\
 & \equiv & \mathcal{L}_{n-1}\left[f\right]\left(t_{0},\dots,t_{n};\mu,z\right)\nonumber 
\end{eqnarray}
where the conditions $f_{n<0}=0$ are implied. The solution of this
equation should be sought in the following form 

\begin{equation}
f_{n}=\bar{f}_{n}\left(t_{2},t_{3},\dots;\mu\right)+\tilde{f}_{n}\left(t_{0},t_{1},\dots;\mu\right)\label{eq:fnDecomp}
\end{equation}
where $\tilde{f}_{n}$ and $\overline{f}_{n}$ are chosen such to
satisfy, respectively, the following two equations:

\begin{equation}
\frac{\partial\tilde{f}_{n}}{\partial t_{0}}-\frac{\partial}{\partial\mu}D\left(\mu\right)\left(1-\mu^{2}\right)\frac{\partial\tilde{f}_{n}}{\partial\mu}=\mathcal{L}_{n-1}\left[\tilde{f}\right]\left(t_{0},\dots,t_{n};\mu,z\right)\label{eq:ftilda}
\end{equation}
and

\begin{equation}
-\frac{\partial}{\partial\mu}D\left(\mu\right)\left(1-\mu^{2}\right)\frac{\partial\bar{f}_{n}}{\partial\mu}=\mathcal{L}_{n-1}\left[\bar{f}\right]\left(t_{2},\dots,t_{n};\mu,z\right)\label{eq:fbar}
\end{equation}
The solution for $\tilde{f}_{n}$ is as follows

\begin{equation}
\tilde{f}_{n}=\sum_{k=1}^{\infty}C_{k}^{(n)}\left(t_{0}\right)e^{-\lambda_{k}t_{0}}\psi_{k}\left(\mu\right)\label{eq:f1tilde}
\end{equation}
and it can be evaluated for arbitrary $n$ by expanding both sides
of eq.(\ref{eq:ftilda}) in a series of eigenfunctions of the diffusion
operator on its l.h.s.: 

\[
-\frac{\partial}{\partial\mu}D\left(\mu\right)\left(1-\mu^{2}\right)\frac{\partial\psi_{k}}{\partial\mu}=\lambda_{k}\psi_{k},
\]
For $D=1$, for example, $\psi_{k}$ are the Legendre polynomials
with $\lambda_{k}=k\left(k+1\right)$, $k=0,1,\dots$. The time dependent
coefficients $C_{k}^{\left(n\right)}$ are determined by the initial
values of $\tilde{f}_{n}$ (anisotropic part of the initial CR distribution)
and the r.h.s. of eq.(\ref{eq:ftilda}), that depends on $\tilde{f}_{n-1}$,
obtained at the preceding step. It is seen, however, that $\tilde{f}_{n}$
exponentially decay in time for $t\gtrsim1$ and we may ignore them%
\footnote{In fact we must do so, as our asymptotic method has a power accuracy
in $\varepsilon\ll1$, but not the exponential accuracy. %
} as we are primarily interested in evolving the system over times
$t\gtrsim\varepsilon^{-2}\gg1$ and even longer. Starting from $n=0$
and using eq.(\ref{eq:fnGen}), for the slowly varying part of $f$
we have 

\begin{equation}
\frac{\partial f_{0}}{\partial t_{0}}=0.\label{eq:df0dt0}
\end{equation}
The solubility condition for $f_{1}$ (obtained by integrating both
sides of eq.{[}\ref{eq:fnGen}{]} in $\mu$) also gives a trivial
result

\begin{equation}
\frac{\partial f_{0}}{\partial t_{1}}=0,\label{eq:df0dt1}
\end{equation}
so the last two conditions are consistent with the suggested decomposition
in eq.(\ref{eq:fnDecomp}), since from eq.(\ref{eq:fbar}) with $n=1$
we have

\begin{equation}
\bar{f}_{1}=-\frac{1}{2}W\frac{\partial f_{0}}{\partial z}\label{eq:f1bar}
\end{equation}
and, thus both $\bar{f}_{0}$ and $\bar{f}_{1}$ are, indeed, independent
of $t_{0}$ and $t_{1}$. We have introduced the function $W\left(\mu\right)$
here by the following two relations

\begin{equation}
\frac{\partial W}{\partial\mu}=\frac{1}{D},\;\;\;\left\langle W\right\rangle =0.\label{eq:Wdef}
\end{equation}
The solubility condition for $f_{2}$ yields the nontrivial and well-known
(e.g., \citealp{Jokipii66}) result, which is actually the leading
term of the $\partial f_{0}/\partial t$ expansion in $\varepsilon\ll1$

\begin{equation}
\frac{\partial f_{0}}{\partial t_{2}}=\frac{1}{4}\left(\frac{\partial}{\partial z}+\sigma\right)\kappa\frac{\partial f_{0}}{\partial z},\label{eq:df0dt2}
\end{equation}
where

\[
\kappa=\left\langle \frac{\left(1-\mu^{2}\right)}{D}\right\rangle .
\]
The solubility conditions for $f_{3}$, $f_{4},...$ will generate
the higher order terms of our expansion which, after some algebra,
can be manipulated into the following expressions for the third and
fourth orders of approximation

\begin{eqnarray}
\frac{\partial f_{0}}{\partial t_{3}} & = & -\frac{1}{4}\left(\frac{\partial}{\partial z}+\sigma\right)\left(\frac{\partial}{\partial z}+\frac{\sigma}{2}\right)\left\langle \mu W^{2}\right\rangle \frac{\partial f_{0}}{\partial z}\label{eq:df0dt3}\\
\frac{\partial f_{0}}{\partial t_{4}} & = & \frac{1}{8}\left(\frac{\partial}{\partial z}+\sigma\right)\times\nonumber \\
 &  & \left\{ \left(\frac{\partial}{\partial z}+\frac{\sigma}{2}\right)^{2}\left\langle W^{2}\left(U^{\prime}-\kappa\right)\right\rangle +\frac{1}{2}\left(\frac{\partial}{\partial z}+\sigma\right)\frac{\partial}{\partial z}\left\langle \frac{\left[\kappa\left(1-\mu\right)+U\right]^{2}}{D\left(1-\mu^{2}\right)}\right\rangle \right\} \frac{\partial f_{0}}{\partial z}.\label{eq:df0dt4}
\end{eqnarray}
We have denoted 

\[
U\equiv\intop_{1}^{\mu}\frac{1-\mu^{2}}{D}d\mu,
\]
and $U^{\prime}=\partial U/\partial\mu$. The pitch-angle diffusion
coefficient $D\left(\mu\right)$ and magnetic focusing $\sigma$ are
considered $z$- independent for simplicity, a limitation that can
be easily relaxed by re-arranging the operators containing $\partial/\partial z$
in eq.(\ref{eq:df0dt4}). We can proceed to higher orders of approximation
\emph{ad infinitum} since terms containing $\left\langle \left(1-\mu^{2}\right)\partial f_{n}/\partial\mu\right\rangle $
can be expressed through $f_{n-1},\; f_{n-2},...$. According to eqs.(\ref{eq:df0dt0}-\ref{eq:df0dt1}),
of interest is the evolution of $f_{0}$ on the time scales $t_{2}\gtrsim1$
or $t\gtrsim\varepsilon^{-2}$ so, as we already mentioned, the contributions
of $\tilde{f}_{n}\left(\mu\right)$ to all the solubility conditions,
similar to those given by eqs.(\ref{eq:df0dt2}-\ref{eq:df0dt4}),
have to be dropped (as they become exponentially small) and only $\bar{f}_{n}\left(\mu\right)$-
contributions should be retained. Using eqs.(\ref{eq:df0dt0}-\ref{eq:df0dt1},\ref{eq:df0dt2}-\ref{eq:df0dt4})
to form the combinations $\varepsilon^{n}\partial^{n}f_{0}/\partial t_{n}$
and summing up both sides, on the l.h.s. of the resulting equation
we simply obtain $\partial f_{0}/\partial t$ (see eq.{[}\ref{eq:TimeDerExpan}{]}).
Therefore, the evolution of $f_{0}$ up to the fourth order in $\varepsilon$
takes the following form

\begin{equation}
\frac{\partial f_{0}}{\partial t}=\frac{\varepsilon^{2}}{4}\partial_{z}^{\prime}\left\{ \kappa-\varepsilon\partial_{z}^{\prime\prime}\left\langle \mu W^{2}\right\rangle -\frac{\varepsilon^{2}}{2}\left[K_{1}\left(\partial_{z}^{\prime\prime}\right)^{2}-K_{2}\partial_{z}^{\prime}\partial_{z}\right]\right\} \frac{\partial f_{0}}{\partial z}\label{eq:Master4}
\end{equation}
where $\partial_{z}^{\prime}=\partial_{z}+\sigma$, $\partial_{z}^{\prime\prime}=\partial_{z}+\sigma/2$,
and

\begin{equation}
K_{1}=\left\langle W^{2}\left(\kappa-U^{\prime}\right)\right\rangle ,\;\;\; K_{2}=\frac{1}{2}\left\langle \frac{\left[\kappa\left(1-\mu\right)+U\right]^{2}}{D\left(1-\mu^{2}\right)}\right\rangle \label{eq:Kdef}
\end{equation}

The above algorithm allows one to obtain the master equation to arbitrary
order in $\varepsilon$. By construction, in no order of approximation
will higher time derivatives emerge, as has been devised by Chapman
and Enskog. We have truncated this process at the fourth order, $\varepsilon^{4}$.
As we show in the next section, this is the lowest order required
to relate the above result to the telegraph equation. It also gives
the first non-vanishing correction to the standard CR diffusion model
in an important case $\left\langle \mu W^{2}\right\rangle =0$, which
is fulfilled, in particular, for $D\left(-\mu\right)=D\left(\mu\right)$.
Higher order terms can be calculated at the expense of a more involved
algebra, but we argue below that such calculations would not change
the results significantly.

\section{Comparison with Earlier Results. Recovering Telegraph Term \label{sec:Comparison}}

In contrast to the telegraph equation given by eqs.(\ref{eq:TelegrEqIntro}-\ref{eq:KappaTau}),
that has been derived by \citet{LitvSchlick13} using direct iteration
of eq.(\ref{eq:Boltzmann}) with no explicit ordering of the emerging
terms, eq.(\ref{eq:Master4}) is derived to the $\varepsilon^{4}$-
order of approximation with an $\varepsilon^{n}$ factor labeling
each term. Yet, it has no second order time derivative which is inconsistent
with eq.(\ref{eq:TelegrEqIntro}). Below we demonstrate that eq.(\ref{eq:Master4})
can still be converted to the telegraph form, however, with additional
terms absent from eq.(\ref{eq:TelegrEqIntro}-\ref{eq:KappaTau}).
Although eq.(\ref{eq:Master4}) is obtained by a broadly a applicable
Chapman-Enskog method, its reduction to the telegraph form below is
more restrictive and should be taken with a grain of salt for the
reasons we discuss later. 

Several versions of telegraph equation have been obtained using different
methods but, unfortunately, many of them do not offer clear ordering,
as e.g., eq.{[}\ref{eq:TelegrEqIntro}{]} derived by \citealp{LitvSchlick13}.
In an earlier treatment by \citep{Earl73}, an eigenfunction expansion
was truncated with no transparent assessment of discarded terms. As
we mentioned already, many treatments do not systematically eliminate
short time scales which are irrelevant to the long-time evolution
of the isotropic part of the CR distribution. In principle, this is
acceptable if the reduction scheme is based on an exact solution of
the original equation, to include all required orders of approximation
into the master equation. Such approach, along with a nearly exhaustive
analysis of the previous work has been presented in \citep{SchwadronTelegraph94}.
Their treatment, however, is by necessity limited to a relatively
simple $D\left(\mu\right)$ (i.e., power-law in $\mu$).

To clarify the role of the higher order terms in eq.(\ref{eq:Master4}),
we note that the r.h.s. of this equation represents just the first
three non-vanishing contributions from an infinite asymptotic series
(in $\varepsilon\ll1$) which we would obtain by continuing the reduction
process described in the preceding section. This series may or may
not converge to some linear (integral) operator in $z$. From a practical
standpoint, the maximum order term that needs to be retained is either
the first non-vanishing term, or else it introduces a new property
to the solution, such as symmetry breaking. Precisely the last aspect
has been highlighted by \citet{LitvSchlick13} who used the telegraph
term to calculate the skewness of a CR pulse. From this angle, we
examine the third and the fourth order below separately.

\paragraph{Third order equation.}

To this order eq.(\ref{eq:Master4}) rewrites

\begin{equation}
\frac{\partial f_{0}}{\partial\tau}=V\frac{\partial f_{0}}{\partial z}+\kappa_{1}\frac{\partial^{2}f_{0}}{\partial z^{2}}-\varepsilon\left\langle \mu W^{2}\right\rangle \frac{\partial^{3}f_{0}}{\partial z^{3}}.\label{eq:eps3}
\end{equation}
We have introduced the slow time $\tau=\varepsilon^{2}t/4$, as a
natural time scale for the reduced system, and the following notation

\[
V=\sigma\left(\kappa-\frac{1}{2}\varepsilon\sigma\left\langle \mu W^{2}\right\rangle \right),\;\;\;\kappa_{1}=\kappa-\frac{3}{2}\varepsilon\sigma\left\langle \mu W^{2}\right\rangle .
\]
Note, that using $\tau$ instead of $t$ makes the terms looking lower
by two orders in $\varepsilon$, but describing them in the sequel
we will use their original order, as it stands in eq.(\ref{eq:Master4})
rather than eqs.(\ref{eq:eps3G}) or (\ref{eq:MasterEps4}). 

Eq.(\ref{eq:eps3}) can be solved using a Fourier transform and integral
representations of Airy functions. We consider some basic properties
of this solution using the moments of $f_{0}$. In particular, it
is seen from this equation that the skewness of a CR pulse, propagating
at the bulk speed $V$, arises in this order of approximation. Indeed,
upon a Galilean transform to the reference frame moving with the speed
$-V$, $z\to z^{\prime}=z+V\tau$, the above equation rewrites

\begin{equation}
\frac{\partial f_{0}}{\partial\tau}=\kappa_{1}\frac{\partial^{2}f_{0}}{\partial z^{\prime2}}-\varepsilon\left\langle \mu W^{2}\right\rangle \frac{\partial^{3}f_{0}}{\partial z^{\prime3}}\label{eq:eps3G}
\end{equation}
so the last term generates an antisymmetric component of $f_{0}\left(z^{\prime}\right)$,
even if $f_{0}\left(z^{\prime}\right)$ is an even function of $z^{\prime}$
initially. By normalizing $f_{0}$ to unity

\[
\bar{f}_{0}=\intop_{-\infty}^{\infty}f_{0}dz^{\prime}=1,
\]
and assuming the coefficients in eq.(\ref{eq:eps3G}) to be constant,
for the moments of $f_{0}\left(z^{\prime},t\right)$ 

\[
\overline{z^{\prime n}}=\intop_{-\infty}^{\infty}f_{0}z^{\prime n}dz^{\prime},
\]
we obtain

\[
\frac{d}{d\tau}\overline{z^{\prime}}=0,\;\;\;\frac{d}{d\tau}\overline{z^{\prime2}}=2\kappa_{1},\;\;\;\frac{d}{d\tau}\overline{z^{\prime3}}=6\kappa_{1}\overline{z^{\prime}}+6\varepsilon\left\langle \mu W^{2}\right\rangle .
\]
With no loss of generality we may set $\overline{z^{\prime}}=0$ and,
in addition, $\overline{z^{\prime3}}=0$ at $\tau=0$, so that the
skewness of the CR distribution changes in time as follows

\[
S\equiv\frac{\overline{z^{\prime3}}}{\left(\overline{z^{\prime2}}\right)^{3/2}}=\frac{6\varepsilon\left\langle \mu W^{2}\right\rangle \tau}{\left(\overline{z_{0}^{\prime2}}+2\kappa_{1}\tau\right)^{3/2}}
\]
where '0' at $\overline{z^{\prime2}}$ refers to its value at $\tau=0$.
The skewness remains small and its maximum 

\[
S_{{\rm max}}=\frac{2\varepsilon\left\langle \mu W^{2}\right\rangle }{\kappa_{1}\sqrt{3\overline{z_{0}^{\prime2}}}}
\]
is achieved at $\tau=\overline{z^{\prime2}}/\kappa_{1}$. Not surprisingly,
the skewness increases with decreasing initial spatial dispersion
of CRs. Indeed, according to eq.(\ref{eq:BoltzmannRescaled}), a narrow
$f\left(z\right)$ generates strong pitch-angle anisotropy which,
in combination with asymmetric pitch-angle scattering ($\left\langle \mu W^{2}\right\rangle \neq0$),
generates the spatial skewness of the CR pulse. For not too small
$\tau$, the explicit form of the solution of eq.(\ref{eq:eps3G})
may be easily written down by using, e.g., a Fourier transform in
$z^{\prime}$ and the steepest descent estimate of its inversion

\begin{equation}
f\left(z^{\prime},\tau\right)\simeq\frac{C}{\sqrt{\tau}}\exp\left[-\frac{z^{\prime2}}{4\kappa_{1}\tau}\left(1-\frac{\varepsilon z^{\prime}}{2\kappa_{1}^{2}\tau}\left\langle \mu W^{2}\right\rangle \right)\right].\label{eq:feps3}
\end{equation}
It is quite possible, however, that even this small effect does not
occur because of the pitch angle scattering symmetry, that is $\left\langle \mu W^{2}\right\rangle =0$.
In this case the solution remains diffusive and, to obtain corrections
to it and to see where the telegraph term might come from, the next
approximation needs to be considered.

\paragraph{Fourth order equation. The telegraph term.}

In the absence of $\varepsilon^{3}$ terms, that is when $\left\langle \mu W^{2}\right\rangle =0$,
eq.(\ref{eq:Master4}) takes the following form

\begin{equation}
\frac{\partial f_{0}}{\partial\tau}+\varepsilon^{2}\left(K_{1}-K_{2}\right)\left(\sigma+\frac{1}{2}\frac{\partial}{\partial z}\right)\frac{\partial^{3}f_{0}}{\partial z^{3}}=\kappa_{2}\frac{\partial^{2}f_{0}}{\partial z^{2}}+V_{2}\frac{\partial f_{0}}{\partial z}+\mathcal{O}\left(\varepsilon^{4}\right)\label{eq:MasterEps4}
\end{equation}

\[
\kappa_{2}=\kappa-\frac{\sigma^{2}\varepsilon^{2}}{2}\left(\frac{5}{4}K_{1}-K_{2}\right),\;\;\; V_{2}=\sigma\left[\kappa-\frac{\sigma^{2}\varepsilon^{2}}{8}K_{1}\right]
\]
This equation, obtained within Chapman-Enskog method, and the telegraph
equation (\ref{eq:TelegrEqIntro}), obtained by a direct iteration
method, differ from each other in the second term on the l.h.s. The
remaining terms of the two equations are equivalent even though not
identical due to the insignificant $\sim\varepsilon^{4}$ corrections
included in the coefficients $\kappa_{2}$ and $V$ on the r.h.s.
of eq.(\ref{eq:MasterEps4}).

To understand how the conflicting terms on the l.h.s. of both equations
are related, we note that within the regular ordering scheme leading
to eq.(\ref{eq:MasterEps4}) the term in question must remain small,
being nominally an $\varepsilon^{4}$-term. However, as the conflicting
terms in both equations are the higher-order derivatives, they may
stick out from their order of approximation if the solution strongly
varies in space and time. In order to preserve the overall solution
integrity in such events a multi- (time)-scale or matched asymptotic
expansion method is normally applied. We will argue that the telegraph
equation approach to the CR transport does not handle this situation
properly, as opposed to the Chapman-Enskog approach. But this is not
to say that the two terms cannot be mapped to each other, when they
are well in the validity range in the above sense. Note that in a
number of other treatments the telegraph term was tacitly handled
as one of the dominant terms. We pointed out that indeed, the term
in question on the l.h.s of eq.(\ref{eq:MasterEps4}) is small only
insofar as the $z$- derivative does not change its order of approximation
due to strong inhomogeneity, whose scale should not be less than the
m.f.p., $\lambda$. So, assuming strong inequality $\partial_{z}\ll\lambda^{-1}$
(or simply $\partial_{z}\sim1$ in our dimensionless variables), we
can express the high order spatial derivatives using a zero order
($\varepsilon\to0$) version of this equation with sufficient accuracy.
The result reads:

\begin{equation}
\frac{\partial f_{0}}{\partial\tau}+\frac{\varepsilon^{2}}{2\kappa^{2}}\left(K_{1}-K_{2}\right)\frac{\partial^{2}f_{0}}{\partial\tau^{2}}=\left(\kappa-\frac{\sigma^{2}\varepsilon^{2}}{8}K_{1}\right)\frac{\partial^{2}f_{0}}{\partial z^{2}}+V_{2}\frac{\partial f_{0}}{\partial z}+\mathcal{O}\left(\varepsilon^{4}\right),\label{eq:TelegrDerived}
\end{equation}
This equation is indeed equivalent to the telegraph equation by its
form, but the equivalence requires not only $\varepsilon\ll1$ but
also smooth variation in $z$ and $\tau$, as not to raise the actual
value of these terms significantly. Under these equivalence conditions,
both the telegraph and the hyperdiffusive transport terms are just
the corrections and may be safely ignored (especially if the $\varepsilon^{3}$
contribution is not empty). On the contrary, when the higher derivatives
strongly enhance these terms, the equations cannot be mapped to each
other and their solutions are disparate. One of them (or even both)
may become less accurate than the underlying leading order (diffusive)
approximation. This is quite common situation in asymptotic expansions
when the form of the next order term should be selected on the ground
of the least possible singularity it introduces into the expansion
(cf. small denominator, secular growth etc. in mechanical problems,
where the higher order approximations, if handled blithely, only aggravate
disagreement with true solutions). The telegraph term correction to
the diffusive approximation appears to come from that variety, as
it generates singular components ($\delta$- and Heaviside functions)
that are not only inconsistent with the strong pitch angle scattering
and resulting spatial diffusion, but with  the scatter-free limit
of the parent differential equation itself. Therefore, the singular
part of the telegraph solution is inherited from the derivation of
telegraph equation. We compare the telegraph and hyperdiffusion type
corrections to the basic diffusive propagation somewhat further in
the next section.

\section{Relation between Telegraph and Hyperdiffusion Approximation\label{sec:Relation-between-Telegraph}}

We start with a relatively minor aspect of the differences between
the two models. As we stated in Sec.\ref{sec:Preliminary-Considerations},
the telegraph coefficient in eq.(\ref{eq:TelegrDerived}) is inconsistent
with some of the earlier derivations. In the simplest case $D=1,$
for example, after proper rescaling of $\tau$ and $z$, it turns
out to be smaller than the term $T$ in eq.(\ref{eq:TelegrEqIntro})
by a factor $11/15$. On the other hand, this is consistent with the
respective result obtained by \citet{Gombosi93,Pauls93} and \citet{SchwadronTelegraph94}.
While the above difference may be considered rather quantitative,
in the general case of $D\left(-\mu\right)\neq D\left(\mu\right)$,
the appropriate equation for describing CR transport is that given
by the lower, $\varepsilon^{3}$- order, not included in eq.(\ref{eq:TelegrEqIntro}).

More importantly, the telegraph version of eq.(\ref{eq:MasterEps4})
given by eq.(\ref{eq:TelegrDerived}) is valid only if the telegraph
term ($\propto\varepsilon^{2}$, fourth order term) remains small
compared to the other terms and the original ordering in eq.(\ref{eq:MasterEps4})
is not violated by strong variations of the solution in space and
time, as we pointed out earlier. We signify this by the ``slow''
time $\tau\sim\varepsilon^{2}t$. In most other treatments $t$ is
used instead, which formally makes the telegraph term in eq.(\ref{eq:TelegrDerived})
appearing as a zero order term. It is not important, of course, whether
the term is labeled by $\varepsilon^{2}$ or not; important is that
it is treated as a subordinate term. Attempts to make it dominant
\emph{a posteriori} violates assumptions that are essential for its
derivation. This point is demonstrated below by repeating a simple
calculation of the CR pulse skewness that we already made earlier
working to $\varepsilon^{3}$ order.

\citet{LitvSchlick13} suggested to study an asymmetry (skewness)
of a CR pulse propagating along the field under the action of magnetic
focusing using the telegraph equation. This and some other characteristics
of the CR pulse, such as the kurtosis, can be easily analyzed using
the primary equation (\ref{eq:MasterEps4}). The calculation of the
pulse skewness essentially repeats the one already done at the $\varepsilon^{3}$
level, where it is generated by asymmetric scattering, $D\left(-\mu\right)\neq D\left(\mu\right)$.
So, transforming eq.(\ref{eq:MasterEps4}) to the reference frame
moving with the speed $-V_{2}$, that is $z\to z^{\prime}=z+V_{2}\tau$,
we obtain

\begin{equation}
\frac{d}{d\tau}\overline{z^{\prime}}=0,\;\;\;\frac{d}{d\tau}\overline{z^{\prime2}}=2\kappa_{2},\;\;\;\frac{d}{d\tau}\overline{z^{\prime3}}=6\varepsilon^{2}\sigma\left(K_{1}-K_{2}\right),\label{eq:Skew4}
\end{equation}
where we have, again, assumed $\overline{z^{\prime}}=0$. The skewness
thus evolves in time as follows

\[
S=\frac{6\varepsilon^{2}\sigma\left(K_{1}-K_{2}\right)\tau}{\left(\overline{z_{0}^{\prime2}}+2\kappa_{2}\tau\right)^{3/2}}
\]
Unless $S\left(\tau\right)$ reaches its maximum very early it is
fairly small. Because $\tau_{{\rm max}}=\overline{z_{0}^{\prime2}}/\kappa_{2}$,
the maximum value

\begin{equation}
S_{{\rm max}}=S\left(\tau_{{\rm max}}\right)=\frac{2\varepsilon^{2}\sigma\left(K_{1}-K_{2}\right)}{\kappa_{2}\sqrt{3\overline{z_{0}^{\prime2}}}},\label{eq:Smax4}
\end{equation}
so that an initially symmetric CR pulse develops significant asymmetry
only if $\overline{z_{0}^{\prime2}}\lesssim\varepsilon^{4}$. This,
however, would require $\tau_{{\rm max}}\sim\varepsilon^{4}$ (or,
equivalently, $t_{{\rm max}}\sim\varepsilon^{2}$), in strong violation
of the requirement $t\gtrsim1$, established in Sec.\ref{sec:Chapman-Enskog-expansion}.
Note that significant pulse asymmetry obtained by \citet{LitvSchlick13}
using the telegraph equation was based on the fundamental solution
to this equation, that is $z_{0}^{\prime}\to0$. We argued in Sec.\ref{sec:Preliminary-Considerations}
that the early propagation phase is not adequately described by the
telegraph equation so that the pulse asymmetry might have been overestimated
in the above paper. We specify the validity range of the telegraph
equation below. 

Starting from a long time regime $\tau>\varepsilon z^{\prime}$, similarly
to the $\varepsilon^{3}$ result given in eq.(\ref{eq:feps3}), from
eq.(\ref{eq:MasterEps4}) we find

\begin{equation}
f\left(z^{\prime},\tau\right)\simeq\frac{C}{\sqrt{\tau}}\exp\left\{ -\frac{z^{\prime2}}{4\kappa_{2}\tau}\left[1-\varepsilon^{2}\frac{\left(K_{1}-K_{2}\right)z^{\prime}}{2\kappa_{2}\tau}\left(\sigma-\frac{z^{\prime}}{4\kappa_{2}\tau}\right)\right]\right\} \label{eq:eps4longtime}
\end{equation}
This propagation regime is not much different from the regular diffusion
(first term in the square bracket), so both the hyperdiffusion and
telegraph models produce similar results, as they are largely equivalent
in this regime. It is worthwhile to write the requirement for the
agreement between the two models in physical units which is simply

\[
vt>z^{\prime}
\]
The point $z=vt$ is close to the cut-off in the telegraph solution,
$z/t=\sqrt{k/T}$ (for $\sigma=0$), eq.(\ref{eq:TelIntroSol}). It
follows then that unless $z^{2}\gg kt$ ( $t\gg T$), the cut-off
strongly changes the overall solution.

The opposite case $\tau<\varepsilon z^{\prime}$, which corresponds
to the initial phase of pulse relaxation, is the key to understanding
the difference between the Chapman-Enskog and the telegraph methods.
In this regime they are \emph{not equivalent}, as the $\varepsilon^{4}$
term in eq.(\ref{eq:MasterEps4}) cannot be neglected to make the
transition to the telegraph equation (\ref{eq:TelegrDerived}). Indeed,
when propagation starts with an infinitely narrow pulse, in the early
phase of its relaxation the higher $z$- derivatives are still too
large for such transformation. A spatially narrow pulse automatically
generates strong pitch angle anisotropy which, in turn, results in
rapid time variation, making the telegraph term also large. It is
this regime where both methods become questionable and, in addition,
their predictions deviate from each other both \emph{quantitatively
and qualitatively}. We need to check first whether they are relevant
to this regime.

The phase $\tau<\varepsilon z^{\prime}$ is well described by the
Chapman-Enskog approach down to $\tau\sim1/\varepsilon^{2}$ ($t_{0}\sim1$,
Sec.\ref{sec:Chapman-Enskog-expansion}). The situation with the telegraph
equation is more complex, as the conversion from the Chapman-Enskog
expansion is invalid, while independent derivations rarely provide
clear ordering. A rigorous derivation in \citep{SchwadronTelegraph94}
requires the same assumptions that we made when transforming the hyperdiffusive
equation into the telegraph equation, that is $\partial_{\tau}\sim\partial_{z}^{2}$
($\epsilon_{\tau}\sim\epsilon_{\lambda}^{2}$ under their nomenclature).
So, the telegraph equation appears to be a subset of the hyperdiffusion
equation valid only under the above ordering. It is likely to break
down in the $\tau<\varepsilon z^{\prime}$ regime, in other words,
near its cut-off. We support this premise by the following considerations. 

Recently, \citet{Effenberger2014} and \citet{LitvEffenSchlick15}
have carried out simulations of the full scattering problem, corresponding
to eq.(\ref{eq:BoltzmannRescaled}). The results deviate from the
telegraph solution precisely at the early phase of the pulse propagation,
when the hyperdiffusion and telegraph models disagree. The two $\delta$-
function pulses with sharp fronts in its solution given by eq.(\ref{eq:TelIntroSol})
are not seen in the simulations. This is understandable, as such features
are inconsistent with the underlying scattering problem. They should
have been smeared out by scattering earlier, since the spatial profile
is shown at five collision times (Fig.2 in \citealp{Effenberger2014}).
Moreover, the $\delta-$ function pulses $\delta\left(z\pm\sqrt{k/T}t\right)$
that are an integral part of the telegraph solution, as they maintain
its normalization, are irrelevant to the primary equation (\ref{eq:PAscatIntro}),
even without collisions. Indeed, if $\mathcal{D}=0$, and the initial
condition is $\delta\left(z\right)$ being constant in $\mu$ for
$-1<\mu<1$, the scatter-free solution is $\delta\left(z-\mu vt\right)$.
Hence, $f_{0}\left(z,t\right)\equiv\intop fd\mu/2=\left(2vt\right)^{-1}H\left(vt-\left|z\right|\right)$.
Therefore, a certain property of the telegraph equation allows the
$\delta\left(z\pm\sqrt{k/T}t\right)$ and sharp front components (although
with modified speed, eq.{[}\ref{eq:TelIntroSol}{]}) to survive multiple
collisions. As these components are inconsistent with the underlying
scattering problem, this property of the equation must have been acquired
during its derivation. Obviously, it is rooted in the hyperbolic (telegraph)
operator $\partial_{t}^{2}-\left(k/T\right)\partial_{z}^{2}$ that
allows the singular profiles to propagate without spreading.

By contrast, the hyperdiffusion equation does not require singular
components but, on the contrary, smears them out. We present an approximate
solution of eq.(\ref{eq:MasterEps4}) after neglecting magnetic focusing
in $\varepsilon^{4}$- order terms in eq.(\ref{eq:MasterEps4}). We
also neglect the regular diffusion compared to the hyperdiffusion,
which is acceptable during an early phase of pulse relaxation, $\tau<\varepsilon z^{\prime}$.
Finally, we assume $z^{\prime}>0$, as the solution is an even function
of $z^{\prime}$. The asymptotic result is as follows:

\begin{equation}
f=\frac{2}{3}\sqrt{\frac{2}{\pi}}\left(4h\tau\right)^{-1/6}z^{\prime-1/3}\exp\left(-\frac{3}{8}4^{-1/3}\frac{z^{\prime4/3}}{h^{1/3}\tau^{1/3}}\right)\cos\left(\frac{3^{3/2}}{8}4^{-1/3}\frac{z^{\prime4/3}}{h^{1/3}\tau^{1/3}}\right).\label{eq:HypDifSmallTau}
\end{equation}
We have denoted the hyperdiffusion constant $h=\varepsilon^{2}\left(K_{1}-K_{2}\right)/2$.
More about this result and further discussion of the two conflicting
approaches can be found in Appendix. We see that there is a considerable
slow down of the CR spreading compared to the conventional diffusion,
$z^{\prime2}\propto t^{1/2}$, that embodies a sub-diffusive propagation,
$z^{\prime2}\propto t^{1/4}$. This ameliorates the problem of acasual
propagation in diffusion regime, yet no sharp fronts or spikes develop.
By contrast, the telegraph solution to the causality problem is to
cut off the solution beyond certain distance ($\left|z\right|>\sqrt{k/T}t$),
thus introducing an unphysical singularity. The immediately arising
normalization problem is then ``solved'' by adding an even stronger
singularity in form of two $\delta$ functions at the cut-off points.

\section{Summary and Conclusions\label{sec:Summary-and-Conclusions}}

Using the Chapman-Enskog method, we have extended the CR transport
equation with magnetic focusing to the fourth order in a small parameter
$\varepsilon=\lambda/l$ (CR mean free path to the characteristic
scale of the problem). This analysis clarifies the nature of the telegraph
transport equation, widely publicized in the literature as a promising
alternative to diffusive propagation models. We have shown that the
telegraph extension ($\propto\partial^{2}f_{0}/\partial t^{2}$) of
the diffusion equation can be mapped from the (small) hyper-diffusive
term ($\propto\partial^{4}f_{0}/\partial z^{4}$) of the regular Chapman-Enskog
expansion, but the telegraph term, originating from an $\varepsilon^{4}$
term of the expansion, must remain subordinate to the main, diffusive
transport and magnetic focusing (if present) contributions. This condition
is met after $\nu t>z/\lambda$ collision times for an initially narrow
($\Delta z<\lambda$) CR distribution. Another important limitation
of the telegraph equation is that, by contrast to the Chapman-Enskog
equation and the original pitch-angle scattering equation, it is not
self-contained requiring an initial condition for also $\partial f_{0}/\partial t$,
which needs information about the anisotropic part of the initial
CR distribution. Furthermore, an attempt to proceed to higher orders
in $\varepsilon$ introduces progressively shorter time scales associated
with ``ghost'' terms reflecting quick relaxation of initial anisotropy
or strong spatial inhomogeneity. By contrast, the classic Chapman-Enskog
method is devised to eliminate short time scales, irrelevant to the
evolution of the isotropic part of CR distribution $f_{0}$, which
accurately describes this evolution after a few collision times, $\nu t>1$. 

We have derived the CR transport equation for an arbitrary pitch-angle
scattering coefficient $D\left(\mu\right)$. This form of transport
equation, (\ref{eq:Master4}), is suitable for describing CR acceleration
and escape problems where the phenomenon of self-confinement ($D$
is a functional of $f$, $D=D\left[f;\mu,t\right]$) is critical,
e.g. \citep{PtuskinNLDIFF08,MDS10PPCF,MetalEsc13,Fujita11}. Accounting
for magnetic focusing effects is required, e.g., for describing particle
acceleration in CR-modified shocks with an oblique magnetic field.
In this case the field increases towards the shocks due to the pressure
exerted by the accelerated CRs on the flow, thus producing a mirror
effect. The particle drift velocity along the field associated with
the mirror effect is (e.g., eq.{[}\ref{eq:eps3}{]}) $V\sim\kappa/l_{B}$
which, for the magnetic field variation scale being of the order of
the shock precursor scale $\kappa/U_{{\rm sh}}$ and strong shock
modification, almost automatically becomes comparable with the shock
velocity $U_{{\rm sh}}$. This additional bulk motion of the accelerated
CR (directed towards the shock) will affect their spectrum and acceleration
time. 

Furthermore, as the CR scattering in such environments (i.e., supernova
remnant {[}SNR{]} shocks) must be self-sustained by virtue of instabilities
of the CR distribution (see, e.g., \citet{BykBrandMalk13,Bell14Br}
for the recent reviews), the above magnetic drift needs to be included
in the CR stability analysis. It should be noted, however, that the
results obtained in the present paper formally require a magnetic
field $B_{0}$ that does not strongly change over the gyro-radius
of energetic particles. This is not to be expected in SNR shocks,
especially if strong, CR current- and pressure-driven instabilities
generate fields with $\delta B>B_{0}$. However, one may use the shock
normal direction as the polar axis to calculate the pitch angle diffusion
coefficient $D\left(\mu\right)$, needed for the description of the
CR spatial transport. Also, such treatment will require a description
of the gyro-motion and averaging by computing particle orbits beyond
the standard quasi-linear description \citep{MD06}, implied throughout
this paper.

In conclusion, by comparison with the telegraph equation, the classic
Chapman-Enskog hyper-diffusion equation consistently describes the
long-term CR propagation in a self-contained, order controlled fashion.
Further improvement of the CR diffusion models should probably address
the anisotropic component of the CR distribution. There are situations,
such as ultra-high energy CR propagation, where the mean free path
grows too long with the energy as to make the diffusive approach irrelevant
and a rectilinear transport to dominate \citep{AloisBerezSuperLum09}
(cf. Levi flight regime described in aforementioned study by \citealp{MD06}).
Another interesting example is a sharp angular anisotropy $\sim10^{\circ}$
in CR arrival directions discovered by MILAGRO observatory \citep{Milagro08PRL}
and a number of other instruments, e.g. \citep{IceAnisFull11,ARGO13,Desiati14,HAWC_Anis14}.
Unless this anisotropy is of a very local origin (such as heliosphere,
\citep{LazarMilagro10,DruryAnisotr13}, it poses a real challenge
to CR propagation models and clearly cannot be addressed within the
diffusive approaches discussed in this paper \citep{DruryAharMILAGRO08,Milagro10,Giacinti12,Ahlers14,2015arXiv150301466M}.
On the other hand, when the diffusive transport model is well within
its validity range (weakly anisotropic spatially smooth CR distributions)
neither the telegraph nor the hyper-diffusive term (both$\sim\varepsilon^{4}$)
is essential to the CR transport and can be neglected.

\acknowledgements{}

Useful and stimulating comments of the Referee are gratefully acknowledged.
The work of MM was supported by the NASA ATP-program under the Grant
NNX14AH36G. He is also indebted to the University of Maryland for
hospitality and partial support during this work. Partial support
from US DoE is also appreciated.

\appendix{}

\section{Appendix\label{sec:Appendix}}

To derive the result given by eq.(\ref{eq:HypDifSmallTau}), we rewrite
eq.(\ref{eq:MasterEps4}) for a simple case of weak focusing ($\sigma f<\partial_{z}f$)
and using the Galilean transform to the frame moving at the magnetic
drift velocity $V_{2},$ $z^{\prime}=z+V_{2}\tau$:

\begin{equation}
\frac{\partial f}{\partial\tau}=\kappa_{2}\frac{\partial^{2}f}{\partial z^{\prime2}}-h\frac{\partial^{4}f}{\partial z^{\prime4}},\label{eq:AppendDiffHyperdiff}
\end{equation}
where $h=\varepsilon^{2}\left(K_{1}-K_{2}\right)/2$ and $\kappa_{2}$
is the same as in eq.(\ref{eq:MasterEps4}). The fundamental solution
of eq.(\ref{eq:AppendDiffHyperdiff}) can be written as an inversion
of the Fourier image $f_{k}\left(\tau\right)$, assuming that $f_{k}\left(0\right)=1/2\pi$.
As we are going to find the Fourier inversion using asymptotic methods,
we write an arbitrary constant $C$ instead of this value:

\begin{equation}
f\left(\tau,z^{\prime}\right)=C\intop_{-\infty}^{\infty}e^{ikz^{\prime}-k^{2}\left(\kappa_{2}+hk^{2}\right)\tau}dk.\label{eq:ApGenSol}
\end{equation}
We will determine the normalization constant $C$ shortly from the
requirement $f\to\delta\left(z^{\prime}\right),\;\tau\to0$. It is
convenient to introduce the following notation: 
\[
\zeta^{4}=4h\tau k^{4},\;\;\xi=z^{\prime}\left(4h\tau\right)^{-1/4}>0.
\]
We may limit our consideration to the $\xi\ge0$ half-space, because
the solution is an even function of $z^{\prime}$. Focusing on a short
time asymptotic regime $\tau<\varepsilon z^{\prime},$ which is opposite
to the case considered earlier (eq.{[}\ref{eq:eps4longtime}{]}) we
neglect the diffusive term $\sim k^{2}$ in the exponent (as hyperdiffusion
dominates) and rewrite eq.(\ref{eq:ApGenSol}) as follows

\begin{equation}
f\left(\tau,z^{\prime}\right)=\frac{C}{\left(4h\tau\right)^{1/4}}\intop_{-\infty}^{\infty}e^{i\xi\zeta-\zeta^{4}/4}d\zeta\label{eq:ApApproxInt}
\end{equation}
Note that the general evaluation of the integral in eq.(\ref{eq:ApGenSol}),
with the diffusive term included, is not difficult but more cumbersome.
The phase of the integral here has three saddle points, and the following
two should be on the integration path 

\[
\zeta_{\pm}=i\xi^{1/3}e^{\pm i\pi/3},
\]
since the integrand reaches its maxima at these points. So, the integration
runs from $-\infty$ through $\zeta_{+}$ to $+i\infty$ then to $\zeta_{-}$
and, finally to $+\infty$. The contributions from the two saddle
points then yield

\begin{equation}
f=\frac{2C\sqrt{\pi/3}}{\left(h\tau\right)^{1/4}\xi^{1/3}}\exp\left(-\frac{3}{8}\xi^{4/3}\right)\cos\left(\frac{3^{3/2}}{8}\xi^{4/3}\right)\label{eq:ApFinSol}
\end{equation}
The normalization $\intop fdz^{\prime}=1$ requires the constant $C=1/\sqrt{3}\pi$,
which only insignificantly deviates from the exact value $C=1/2\pi$
that, in turn, follows from the integral representation of $f$ in
eq.(\ref{eq:ApGenSol}) for $\tau\to0$. Note that we could have replaced
an oscillating exponential tail of this solution by zero beyond the
point $\xi>\left(4\pi\right)^{3/4}3^{-9/8}$. Such modification of
the hyperdiffusive solution would be in the spirit of the telegraph
cut-off at $z/t=\sqrt{k/T}$, however, with an essential difference
of being only a discontinuity in the solution derivative. The unphysical
(oscillatory) behavior beyond the first zero point of the solution
in eq.(\ref{eq:ApFinSol}) results from neglecting the diffusion term
in eq.(\ref{eq:AppendDiffHyperdiff}), asymptotic methods used to
calculate the integral in eq.(\ref{eq:ApApproxInt}), and from lacking
higher order terms, $\sim\varepsilon^{n},\; n>4$. Therefore, the
solution can be improved systematically. In addition, it starts from
a point source which is clearly inconsistent with the main approximation
$\varepsilon\ll1$. A somewhat broader initial profile will not develop
an oscillatory tail, if convolved with the Green's function in eq.(\ref{eq:ApGenSol}).
We will not attempt to improve on this minor aspect of the solution
here, as it becomes only weakly irregular if cut off at its first
zero. 

From the perspective of a general improvement of the asymptotic expansion
considered in this paper, the derivation of eqs.(\ref{eq:Skew4}),
for example, is robust in the following sense. The residual higher
order terms in $\varepsilon\ll1$, if included in eq.(\ref{eq:MasterEps4}),
will not change eqs.(\ref{eq:Skew4}) in any other way than small
corrections to the coefficients $\kappa_{2}$ and $V_{2}$. Indeed,
the higher $z$- derivatives coming from higher order terms, will
vanish from the (first four) moment equations after integrating by
parts. By contrast, continuing the telegraph approach to higher orders
will generate terms with small parameters at higher time derivatives
in all moment equations. These terms will become crucial during the
initial relaxation of the CR distribution. The relaxation is associated
with the CR anisotropy or strong initial inhomogeneity, that is with
large $\tilde{f_{n}}$, Sec.\ref{sec:Chapman-Enskog-expansion}. However,
these decay over a short time $t\lesssim1$. This is the time period
when the telegraph or hyperdiffusive correction is large but its effect
on the subsequent evolution ought to be limited as this time is short.
The hyperdiffusive correction meets this requirement, as we argued
using moment equations. To see whether the same is true for the telegraph
correction, let us rewrite eq.(\ref{eq:TelegrDerived}) using the
``fast'' time, $t=\tau/\varepsilon^{2}$:

\begin{equation}
\left(1+\tau_{{\rm T}}\frac{\partial}{\partial t}\right)\frac{\partial f_{0}}{\partial t}=\frac{\varepsilon^{2}}{4}\kappa\frac{\partial^{2}f_{0}}{\partial z^{2}},\label{eq:AppTelegrRescaled}
\end{equation}
where we denoted $\tau_{{\rm T}}=2\left(K_{1}-K_{2}\right)/\kappa\sim1$
and assumed $\sigma=0$, to make the following simple argument. Namely,
in the limit $\varepsilon\to0$ there are two modes, of which the
first is $f_{0}=f_{0}\left(z\right)$. This is the main diffusion
mode that slowly evolves in time when $0<\varepsilon\ll1$, and, as
we are interested in the evolution over the time scales $t\gtrsim\varepsilon^{-2}$,
the telegraph term becomes $\sim\varepsilon^{4}$ and can be discarded.
The second mode corresponds to a rapid decay of the initial distribution
$\sim{\rm exp}\left(-t/\tau_{{\rm T}}\right)$ which is associated
with the decay of initial anisotropy or strong inhomogeneity. If this
mode is active ($\partial f_{0}/\partial t$$\neq0$ in eq.{[}\ref{eq:AppTelegrRescaled}{]}),
then even the total number of particles $N$ is not conserved automatically.
So, turning to the moments of eq.(\ref{eq:AppTelegrRescaled}) we
need to impose the initial condition, $\partial N/\partial t=0$,
to ensure the particle conservation. This probably means that $\varepsilon\to0$
is a difficult limit for the telegraph reduction scheme. The singular
components in the telegraph solution (\ref{eq:TelIntroSol}) appear
to be primarily associated with the particle conservation problem.
The initial relaxation phase ($t\lesssim1$) perhaps, cannot be adequately
described by the telegraph reduction scheme using an equation for
$f_{0}$ alone, as it does not properly ``average out'' an anisotropic
component $\tilde{f}$, which is large during this period of time.
The telegraph term is therefore to be understood as a ``ghost''
term reflecting rapid decay of such components. It follows that the
rapidly changing part $\tilde{f}$ in the decomposition in eq.(\ref{eq:fnDecomp})
needs to be retained in the short-time analysis along with $f_{0}$.
Otherwise, the telegraph operator generates unphysical $\delta$-
pulses and sharp fronts, just to conserve the number of particles,
as discussed earlier. These considerations are, however, not nearly
complete. Further useful analysis of propagation modes in the context
of telegraph equation can be found in \citep{SchwadronTelegraph94}.

To conclude this Appendix we make yet another argument in disfavor
of the telegraph equation that is partially related to the above considerations.
A consistent asymptotic reduction method must be continuable to infinity
in powers of small $\varepsilon$. The Chapman-Enskog scheme clearly
is. The outcome will be a series of terms $\sim\partial_{z}^{n}f_{0}$
on the r.h.s. of eq.(\ref{eq:Master4}) with just $\partial_{t}f_{0}$
on its l.h.s. To solve the resulting equation, only the initial distribution
$f_{0}\left(0,z\right)$ is needed, as the equation remains \emph{evolutionary}
and (generalized) parabolic, as its pitch-angle diffusion superset
is. The telegraph equation, on the contrary, turns hyperbolic and
non-evolutionary after the reduction from the superset equation. By
continuing the telegraph reduction scheme to higher orders of approximation,
progressively higher \emph{time derivatives} will emerge (along with
higher space derivatives). The resulting equations will thus be non-evolutionary
and a growing set of initial time derivatives $\partial^{n}$$f_{0}$
will then be needed to solve the initial value problem. These data
can be extracted only from the full anisotropic distribution with
recourse to the full (anisotropic) equation. Therefore, the telegraph
equation is not self-contained and cannot be improved systematically.
Any attempts to improve it will introduce shorter and shorter time
scales that would require a return to the full anisotropic description.

\bibliographystyle{apj}
\bibliography{}

\begin{thebibliography}{40}
\expandafter\ifx\csname natexlab\endcsname\relax\def\natexlab#1{#1}\fi

\bibitem[{Abbasi {et~al.}({2011})Abbasi, Abdou, Abu-Zayyad, Adams, Aguilar,
  Ahlers, Altmann, Andeen, Auffenberg, Bai, Baker, Barwick, Bay, Alba, Beattie,
  Beatty, Bechet, Becker, Becker, Benabderrahmane, BenZvi, Berdermann,
  Berghaus, Berley, Bernardini, Bertrand, Besson, Bindig, Bissok, Blaufuss,
  Blumenthal, Boersma, Bohm, Bose, Boeser, Botner, Brown, Buitink,
  Caballero-Mora, Carson, Chirkin, Christy, Clem, Clevermann, Cohen, Colnard,
  Cowen, D'Agostino, Danninger, Daughhetee, Davis, De~Clercq, Demiroers,
  Denger, Depaepe, Descamps, Desiati, de~Vries-Uiterweerd, DeYoung, Diaz-Velez,
  Dierckxsens, Dreyer, Dumm, Ehrlich, Eisch, Ellsworth, Engdegard, Euler,
  Evenson, Fadiran, Fazely, Fedynitch, Feintzeig, Feusels, Filimonov, Finley,
  Fischer-Wasels, Foerster, Fox, Franckowiak, Franke, Gaisser, Gallagher,
  Gerhardt, Gladstone, Gluesenkamp, Goldschmidt, Goodman, Gora, Grant, Griesel,
  Gross, Grullon, Gurtner, Ha, Hajismail, Hallgren, Halzen, Han, Hanson,
  Heinen, Helbing, Herquet, Hickford, Hill, Hoffman, Homeier, Hoshina, Hubert,
  Huelsnitz, Huelss, Hulth, Hultqvist, Hussain, Ishihara, Jacobsen, Japaridze,
  Johansson, Joseph, Kampert, Kappes, Karg, Karle, Kenny, Kiryluk, Kislat,
  Klein, Koehne, Kohnen, Kolanoski, Koepke, Kopper, Koskinen, Kowalski,
  Kowarik, Krasberg, Krings, Kroll, Kurahashi, Kuwabara, Labare, Lafebre,
  Laihem, Landsman, Larson, Lauer, Luenemann, Madajczyk, Madsen, Majumdar,
  Marotta, Maruyama, Mase, Matis, Meagher, Merck, Meszaros, Meures, Middell,
  Milke, Miller, Montaruli, Morse, Movit, Nahnhauer, Nam, Naumann, Niessen,
  Nygren, Odrowski, Olivas, Olivo, O'Murchadha, Ono, Panknin, Paul, de~los
  Heros, Petrovic, Piegsa, Pieloth, Porrata, Posselt, Price, Price, Przybylski,
  Rawlins, Redl, Resconi, Rhode, Ribordy, Rizzo, Rodrigues, Roth, Rothmaier,
  Rott, Ruhe, Rutledge, Ruzybayev, Ryckbosch, Sander, Santander, Sarkar,
  Schatto, Schmidt, Schoenwald, Schukraft, Schultes, Schulz, Schunck, Seckel,
  Semburg, Seo, Sestayo, Seunarine, Silvestri, Slipak, Spiczak, Spiering,
  Stamatikos, Stanev, Stephens, Stezelberger, Stokstad, Stoessl, Stoyanov,
  Strahler, Straszheim, Stuer, Sullivan, Swillens, Taavola, Taboada, Tamburro,
  Tepe, Ter-Antonyan, Tilav, Toale, Toscano, Tosi, Turcan, van Eijndhoven,
  Vandenbroucke, Van~Overloop, van Santen, Vehring, Voge, Walck, Waldenmaier,
  Wallraff, Walter, Weaver, Wendt, Westerhoff, Whitehorn, Wiebe, Wiebusch,
  Williams, Wischnewski, Wissing, Wolf, Wood, Woschnagg, Xu, Xu, Yodh, Yoshida,
  Zarzhitsky, Zoll, \& Collaboration}]{IceAnisFull11}
Abbasi, R., Abdou, Y., Abu-Zayyad, T., Adams, J., Aguilar, J.~A., Ahlers, M.,
  Altmann, D., Andeen, K., Auffenberg, J., Bai, X., Baker, M., Barwick, S.~W.,
  Bay, R., Alba, J. L.~B., Beattie, K., Beatty, J.~J., Bechet, S., Becker,
  J.~K., Becker, K.~H., Benabderrahmane, M.~L., BenZvi, S., Berdermann, J.,
  Berghaus, P., Berley, D., Bernardini, E., Bertrand, D., Besson, D.~Z.,
  Bindig, D., Bissok, M., Blaufuss, E., Blumenthal, J., Boersma, D.~J., Bohm,
  C., Bose, D., Boeser, S., Botner, O., Brown, A.~M., Buitink, S.,
  Caballero-Mora, K.~S., Carson, M., Chirkin, D., Christy, B., Clem, J.,
  Clevermann, F., Cohen, S., Colnard, C., Cowen, D.~F., D'Agostino, M.~V.,
  Danninger, M., Daughhetee, J., Davis, J.~C., De~Clercq, C., Demiroers, L.,
  Denger, T., Depaepe, O., Descamps, F., Desiati, P., de~Vries-Uiterweerd, G.,
  DeYoung, T., Diaz-Velez, J.~C., Dierckxsens, M., Dreyer, J., Dumm, J.~P.,
  Ehrlich, R., Eisch, J., Ellsworth, R.~W., Engdegard, O., Euler, S., Evenson,
  P.~A., Fadiran, O., Fazely, A.~R., Fedynitch, A., Feintzeig, J., Feusels, T.,
  Filimonov, K., Finley, C., Fischer-Wasels, T., Foerster, M.~M., Fox, B.~D.,
  Franckowiak, A., Franke, R., Gaisser, T.~K., Gallagher, J., Gerhardt, L.,
  Gladstone, L., Gluesenkamp, T., Goldschmidt, A., Goodman, J.~A., Gora, D.,
  Grant, D., Griesel, T., Gross, A., Grullon, S., Gurtner, M., Ha, C.,
  Hajismail, A., Hallgren, A., Halzen, F., Han, K., Hanson, K., Heinen, D.,
  Helbing, K., Herquet, P., Hickford, S., Hill, G.~C., Hoffman, K.~D., Homeier,
  A., Hoshina, K., Hubert, D., Huelsnitz, W., Huelss, J.~P., Hulth, P.~O.,
  Hultqvist, K., Hussain, S., Ishihara, A., Jacobsen, J., Japaridze, G.~S.,
  Johansson, H., Joseph, J.~M., Kampert, K.~H., Kappes, A., Karg, T., Karle,
  A., Kenny, P., Kiryluk, J., Kislat, F., Klein, S.~R., Koehne, J.~H., Kohnen,
  G., Kolanoski, H., Koepke, L., Kopper, S., Koskinen, D.~J., Kowalski, M.,
  Kowarik, T., Krasberg, M., Krings, T., Kroll, G., Kurahashi, N., Kuwabara,
  T., Labare, M., Lafebre, S., Laihem, K., Landsman, H., Larson, M.~J., Lauer,
  R., Luenemann, J., Madajczyk, B., Madsen, J., Majumdar, P., Marotta, A.,
  Maruyama, R., Mase, K., Matis, H.~S., Meagher, K., Merck, M., Meszaros, P.,
  Meures, T., Middell, E., Milke, N., Miller, J., Montaruli, T., Morse, R.,
  Movit, S.~M., Nahnhauer, R., Nam, J.~W., Naumann, U., Niessen, P., Nygren,
  D.~R., Odrowski, S., Olivas, A., Olivo, M., O'Murchadha, A., Ono, M.,
  Panknin, S., Paul, L., de~los Heros, C.~P., Petrovic, J., Piegsa, A.,
  Pieloth, D., Porrata, R., Posselt, J., Price, C.~C., Price, P.~B.,
  Przybylski, G.~T., Rawlins, K., Redl, P., Resconi, E., Rhode, W., Ribordy,
  M., Rizzo, A., Rodrigues, J.~P., Roth, P., Rothmaier, F., Rott, C., Ruhe, T.,
  Rutledge, D., Ruzybayev, B., Ryckbosch, D., Sander, H.~G., Santander, M.,
  Sarkar, S., Schatto, K., Schmidt, T., Schoenwald, A., Schukraft, A.,
  Schultes, A., Schulz, O., Schunck, M., Seckel, D., Semburg, B., Seo, S.~H.,
  Sestayo, Y., Seunarine, S., Silvestri, A., Slipak, A., Spiczak, G.~M.,
  Spiering, C., Stamatikos, M., Stanev, T., Stephens, G., Stezelberger, T.,
  Stokstad, R.~G., Stoessl, A., Stoyanov, S., Strahler, E.~A., Straszheim, T.,
  Stuer, M., Sullivan, G.~W., Swillens, Q., Taavola, H., Taboada, I., Tamburro,
  A., Tepe, A., Ter-Antonyan, S., Tilav, S., Toale, P.~A., Toscano, S., Tosi,
  D., Turcan, D., van Eijndhoven, N., Vandenbroucke, J., Van~Overloop, A., van
  Santen, J., Vehring, M., Voge, M., Walck, C., Waldenmaier, T., Wallraff, M.,
  Walter, M., Weaver, C., Wendt, C., Westerhoff, S., Whitehorn, N., Wiebe, K.,
  Wiebusch, C.~H., Williams, D.~R., Wischnewski, R., Wissing, H., Wolf, M.,
  Wood, T.~R., Woschnagg, K., Xu, C., Xu, X.~W., Yodh, G., Yoshida, S.,
  Zarzhitsky, P., Zoll, M., \& Collaboration, I. {2011}, {ASTROPHYSICAL
  JOURNAL}, {740}

\bibitem[{{Abdo} {et~al.}(2008){Abdo}, {Allen}, {Aune}, {Berley}, {Blaufuss},
  {Casanova}, {Chen}, {Dingus}, {Ellsworth}, {Fleysher}, {Fleysher},
  {Gonzalez}, {Goodman}, {Hoffman}, {H{\"u}ntemeyer}, {Kolterman}, {Lansdell},
  {Linnemann}, {McEnery}, {Mincer}, {Nemethy}, {Noyes}, {Pretz}, {Ryan},
  {Parkinson}, {Shoup}, {Sinnis}, {Smith}, {Sullivan}, {Vasileiou}, {Walker},
  {Williams}, \& {Yodh}}]{Milagro08PRL}
{Abdo}, A.~A., {Allen}, B., {Aune}, T., {Berley}, D., {Blaufuss}, E.,
  {Casanova}, S., {Chen}, C., {Dingus}, B.~L., {Ellsworth}, R.~W., {Fleysher},
  L., {Fleysher}, R., {Gonzalez}, M.~M., {Goodman}, J.~A., {Hoffman}, C.~M.,
  {H{\"u}ntemeyer}, P.~H., {Kolterman}, B.~E., {Lansdell}, C.~P., {Linnemann},
  J.~T., {McEnery}, J.~E., {Mincer}, A.~I., {Nemethy}, P., {Noyes}, D.,
  {Pretz}, J., {Ryan}, J.~M., {Parkinson}, P.~M.~S., {Shoup}, A., {Sinnis}, G.,
  {Smith}, A.~J., {Sullivan}, G.~W., {Vasileiou}, V., {Walker}, G.~P.,
  {Williams}, D.~A., \& {Yodh}, G.~B. 2008, Physical Review Letters, 101,
  221101

\bibitem[{{Abeysekara} {et~al.}(2014){Abeysekara}, {Alfaro}, {Alvarez},
  {{\'A}lvarez}, {Arceo}, {Arteaga-Vel{\'a}zquez}, {Ayala Solares}, {Barber},
  {Baughman}, {Bautista-Elivar}, {Belmont}, {BenZvi}, {Berley}, {Bonilla
  Rosales}, {Braun}, {Caballero-Mora}, {Carrami{\~n}ana}, {Castillo}, {Cotti},
  {Cotzomi}, {de la Fuente}, {De Le{\'o}n}, {DeYoung}, {Diaz Hernandez},
  {D{\'{\i}}az-V{\'e}lez}, {Dingus}, {DuVernois}, {Ellsworth}, {Fiorino},
  {Fraija}, {Galindo}, {Garfias}, {Gonz{\'a}lez}, {Goodman}, {Gussert},
  {Hampel-Arias}, {Harding}, {H{\"u}ntemeyer}, {Hui}, {Imran}, {Iriarte},
  {Karn}, {Kieda}, {Kunde}, {Lara}, {Lauer}, {Lee}, {Lennarz}, {Le{\'o}n
  Vargas}, {Linnemann}, {Longo}, {Luna-Garc{\'{\i}}a}, {Malone}, {Marinelli},
  {Marinelli}, {Martinez}, {Martinez}, {Mart{\'{\i}}nez-Castro}, {Matthews},
  {McEnery}, {Mendoza Torres}, {Miranda-Romagnoli}, {Moreno}, {Mostaf{\'a}},
  {Nellen}, {Newbold}, {Noriega-Papaqui}, {Oceguera-Becerra}, {Patricelli},
  {Pelayo}, {P{\'e}rez-P{\'e}rez}, {Pretz}, {Rivi{\`e}re}, {Rosa-Gonz{\'a}lez},
  {Ruiz-Velasco}, {Ryan}, {Salazar}, {Salesa Greus}, {Sandoval}, {Schneider},
  {Sinnis}, {Smith}, {Sparks Woodle}, {Springer}, {Taboada}, {Toale},
  {Tollefson}, {Torres}, {Ukwatta}, {Villase{\~n}or}, {Weisgarber},
  {Westerhoff}, {Wisher}, {Wood}, {Yodh}, {Younk}, {Zaborov}, {Zepeda}, {Zhou},
  \& {The HAWC Collaboration}}]{HAWC_Anis14}
{Abeysekara}, A.~U., {Alfaro}, R., {Alvarez}, C., {{\'A}lvarez}, J.~D.,
  {Arceo}, R., {Arteaga-Vel{\'a}zquez}, J.~C., {Ayala Solares}, H.~A.,
  {Barber}, A.~S., {Baughman}, B.~M., {Bautista-Elivar}, N., {Belmont}, E.,
  {BenZvi}, S.~Y., {Berley}, D., {Bonilla Rosales}, M., {Braun}, J.,
  {Caballero-Mora}, K.~S., {Carrami{\~n}ana}, A., {Castillo}, M., {Cotti}, U.,
  {Cotzomi}, J., {de la Fuente}, E., {De Le{\'o}n}, C., {DeYoung}, T., {Diaz
  Hernandez}, R., {D{\'{\i}}az-V{\'e}lez}, J.~C., {Dingus}, B.~L., {DuVernois},
  M.~A., {Ellsworth}, R.~W., {Fiorino}, D.~W., {Fraija}, N., {Galindo}, A.,
  {Garfias}, F., {Gonz{\'a}lez}, M.~M., {Goodman}, J.~A., {Gussert}, M.,
  {Hampel-Arias}, Z., {Harding}, J.~P., {H{\"u}ntemeyer}, P., {Hui}, C.~M.,
  {Imran}, A., {Iriarte}, A., {Karn}, P., {Kieda}, D., {Kunde}, G.~J., {Lara},
  A., {Lauer}, R.~J., {Lee}, W.~H., {Lennarz}, D., {Le{\'o}n Vargas}, H.,
  {Linnemann}, J.~T., {Longo}, M., {Luna-Garc{\'{\i}}a}, R., {Malone}, K.,
  {Marinelli}, A., {Marinelli}, S.~S., {Martinez}, H., {Martinez}, O.,
  {Mart{\'{\i}}nez-Castro}, J., {Matthews}, J.~A.~J., {McEnery}, J., {Mendoza
  Torres}, E., {Miranda-Romagnoli}, P., {Moreno}, E., {Mostaf{\'a}}, M.,
  {Nellen}, L., {Newbold}, M., {Noriega-Papaqui}, R., {Oceguera-Becerra}, T.,
  {Patricelli}, B., {Pelayo}, R., {P{\'e}rez-P{\'e}rez}, E.~G., {Pretz}, J.,
  {Rivi{\`e}re}, C., {Rosa-Gonz{\'a}lez}, D., {Ruiz-Velasco}, E., {Ryan}, J.,
  {Salazar}, H., {Salesa Greus}, F., {Sandoval}, A., {Schneider}, M., {Sinnis},
  G., {Smith}, A.~J., {Sparks Woodle}, K., {Springer}, R.~W., {Taboada}, I.,
  {Toale}, P.~A., {Tollefson}, K., {Torres}, I., {Ukwatta}, T.~N.,
  {Villase{\~n}or}, L., {Weisgarber}, T., {Westerhoff}, S., {Wisher}, I.~G.,
  {Wood}, J., {Yodh}, G.~B., {Younk}, P.~W., {Zaborov}, D., {Zepeda}, A.,
  {Zhou}, H., \& {The HAWC Collaboration}. 2014, \apj, 796, 108

\bibitem[{{Ahlers}(2014)}]{Ahlers14}
{Ahlers}, M. 2014, Physical Review Letters, 112, 021101

\bibitem[{{Aloisio} {et~al.}(2009){Aloisio}, {Berezinsky}, \&
  {Gazizov}}]{AloisBerezSuperLum09}
{Aloisio}, R., {Berezinsky}, V., \& {Gazizov}, A. 2009, \apj, 693, 1275

\bibitem[{{Axford}(1965)}]{AxfordTelegr65}
{Axford}, W.~I. 1965, \planss, 13, 1301

\bibitem[{Bartoli {et~al.}(2013)Bartoli, Bernardini, Bi, Bolognino, Branchini,
  Budano, Calabrese~Melcarne, Camarri, Cao, Cardarelli, Catalanotti, Chen,
  Chen, Creti, Cui, Dai, D'Amone, Danzengluobu, De~Mitri, D'Ettorre~Piazzoli,
  Di~Girolamo, Di~Sciascio, Feng, Feng, Feng, Gou, Guo, He, Hu, Hu, Iacovacci,
  Iuppa, Jia, Labaciren, Li, Liguori, Liu, Liu, Liu, Lu, Ma, Mancarella, Mari,
  Marsella, Martello, Mastroianni, Montini, Ning, Panareo, Panico, Perrone,
  Pistilli, Ruggieri, Salvini, Santonico, Sbano, Shen, Sheng, Shi, Surdo, Tan,
  Vallania, Vernetto, Vigorito, Wang, Wu, Wu, Xue, Yan, Yang, Yang, Yao, Yuan,
  Zha, Zhang, Zhang, Zhang, Zhang, Zhaxiciren, Zhaxisangzhu, Zhou, Zhu, Zhu, \&
  Zizzi}]{ARGO13}
Bartoli, B., Bernardini, P., Bi, X.~J., Bolognino, I., Branchini, P., Budano,
  A., Calabrese~Melcarne, A.~K., Camarri, P., Cao, Z., Cardarelli, R.,
  Catalanotti, S., Chen, S.~Z., Chen, T.~L., Creti, P., Cui, S.~W., Dai, B.~Z.,
  D'Amone, A., Danzengluobu, De~Mitri, I., D'Ettorre~Piazzoli, B., Di~Girolamo,
  T., Di~Sciascio, G., Feng, C.~F., Feng, Z., Feng, Z., Gou, Q.~B., Guo, Y.~Q.,
  He, H.~H., Hu, H., Hu, H., Iacovacci, M., Iuppa, R., Jia, H.~Y., Labaciren,
  Li, H.~J., Liguori, G., Liu, C., Liu, J., Liu, M.~Y., Lu, H., Ma, X.~H.,
  Mancarella, G., Mari, S.~M., Marsella, G., Martello, D., Mastroianni, S.,
  Montini, P., Ning, C.~C., Panareo, M., Panico, B., Perrone, L., Pistilli, P.,
  Ruggieri, F., Salvini, P., Santonico, R., Sbano, S.~N., Shen, P.~R., Sheng,
  X.~D., Shi, F., Surdo, A., Tan, Y.~H., Vallania, P., Vernetto, S., Vigorito,
  C., Wang, H., Wu, C.~Y., Wu, H.~R., Xue, L., Yan, Y.~X., Yang, Q.~Y., Yang,
  X.~C., Yao, Z.~G., Yuan, A.~F., Zha, M., Zhang, H.~M., Zhang, L., Zhang,
  X.~Y., Zhang, Y., Zhaxiciren, Zhaxisangzhu, Zhou, X.~X., Zhu, F.~R., Zhu,
  Q.~Q., \& Zizzi, G. 2013, Phys. Rev. D, 88, 082001

\bibitem[{{Bell}(2014)}]{Bell14Br}
{Bell}, A.~R. 2014, Brazilian Journal of Physics

\bibitem[{{Braginskii}(1965)}]{Braginsky65}
{Braginskii}, S.~I. 1965, {Transport Processes in a Plasma}, Vol.~1, 205

\bibitem[{{Bykov} {et~al.}(2013){Bykov}, {Brandenburg}, {Malkov}, \&
  {Osipov}}]{BykBrandMalk13}
{Bykov}, A.~M., {Brandenburg}, A., {Malkov}, M.~A., \& {Osipov}, S.~M. 2013,
  \ssr, 178, 201

\bibitem[{Cercignani(1988)}]{cercignani1988boltzmannequation}
Cercignani, C. 1988, BoltzmannEquation, Applied Mathematical Sciences Series
  (Springer-Verlag)

\bibitem[{{Chapman} \& {Cowling}(1991)}]{ChapmanCowling1991}
{Chapman}, S., \& {Cowling}, T.~G. 1991, {The Mathematical Theory of
  Non-uniform Gases}

\bibitem[{{Desiati}(2014)}]{Desiati14}
{Desiati}, P. 2014, Nuclear Instruments and Methods in Physics Research A, 742,
  199

\bibitem[{{Drury} \& {Aharonian}(2008)}]{DruryAharMILAGRO08}
{Drury}, L.~O.~C., \& {Aharonian}, F.~A. 2008, Astroparticle Physics, 29, 420

\bibitem[{{Earl}(1973)}]{Earl73}
{Earl}, J.~A. 1973, \apj, 180, 227

\bibitem[{{Effenberger} \& {Litvinenko}(2014)}]{Effenberger2014}
{Effenberger}, F., \& {Litvinenko}, Y.~E. 2014, \apj, 783, 15

\bibitem[{{Fujita} {et~al.}(2011){Fujita}, {Takahara}, {Ohira}, \&
  {Iwasaki}}]{Fujita11}
{Fujita}, Y., {Takahara}, F., {Ohira}, Y., \& {Iwasaki}, K. 2011, \mnras, 415,
  3434

\bibitem[{{Giacinti} \& {Sigl}(2012)}]{Giacinti12}
{Giacinti}, G., \& {Sigl}, G. 2012, Physical Review Letters, 109, 071101

\bibitem[{Goldstein(1951)}]{GOLDSTEIN01011951}
Goldstein, S. 1951, The Quarterly Journal of Mechanics and Applied Mathematics,
  4, 129

\bibitem[{{Gombosi} {et~al.}(1993){Gombosi}, {Jokipii}, {Kota}, {Lorencz}, \&
  {Williams}}]{Gombosi93}
{Gombosi}, T.~I., {Jokipii}, J.~R., {Kota}, J., {Lorencz}, K., \& {Williams},
  L.~L. 1993, \apj, 403, 377

\bibitem[{{Gurevich}(1961)}]{Gurevich61RunAway}
{Gurevich}, A.~V. 1961, Soviet Journal of Experimental and Theoretical Physics,
  12, 904

\bibitem[{{Jokipii}(1966)}]{Jokipii66}
{Jokipii}, J.~R. 1966, \apj, 146, 480

\bibitem[{{Kruskal} \& {Bernstein}(1964)}]{KruskalBernstein64}
{Kruskal}, M.~D., \& {Bernstein}, I.~B. 1964, Physics of Fluids, 7, 407

\bibitem[{{Kulsrud}(2005)}]{Kulsrud05}
{Kulsrud}, R.~M. 2005, {Plasma physics for astrophysics}

\bibitem[{{Lazarian} \& {Desiati}(2010)}]{LazarMilagro10}
{Lazarian}, A., \& {Desiati}, P. 2010, \apj, 722, 188

\bibitem[{{Litvinenko} {et~al.}(2015){Litvinenko}, {Effenberger}, \&
  {Schlickeiser}}]{LitvEffenSchlick15}
{Litvinenko}, Y.~E., {Effenberger}, F., \& {Schlickeiser}, R. 2015, ArXiv
  e-prints

\bibitem[{{Litvinenko} \& {Noble}(2013)}]{LitvNoble13}
{Litvinenko}, Y.~E., \& {Noble}, P.~L. 2013, \apj, 765, 31

\bibitem[{{Litvinenko} \& {Schlickeiser}(2013)}]{LitvSchlick13}
{Litvinenko}, Y.~E., \& {Schlickeiser}, R. 2013, \aap, 554, A59

\bibitem[{{Malkov}(2015)}]{2015arXiv150301466M}
{Malkov}, M.~A. 2015, ArXiv e-prints

\bibitem[{{Malkov} \& {Diamond}(2006)}]{MD06}
{Malkov}, M.~A., \& {Diamond}, P.~H. 2006, \apj, 642, 244

\bibitem[{{Malkov} {et~al.}(2010{\natexlab{a}}){Malkov}, {Diamond},
  {O'C.~Drury}, \& {Sagdeev}}]{Milagro10}
{Malkov}, M.~A., {Diamond}, P.~H., {O'C.~Drury}, L., \& {Sagdeev}, R.~Z.
  2010{\natexlab{a}}, \apj, 721, 750

\bibitem[{{Malkov} {et~al.}(2010{\natexlab{b}}){Malkov}, {Diamond}, \&
  {Sagdeev}}]{MDS10PPCF}
{Malkov}, M.~A., {Diamond}, P.~H., \& {Sagdeev}, R.~Z. 2010{\natexlab{b}},
  Plasma Physics and Controlled Fusion, 52, 124006

\bibitem[{{Malkov} {et~al.}(2013){Malkov}, {Diamond}, {Sagdeev}, {Aharonian},
  \& {Moskalenko}}]{MetalEsc13}
{Malkov}, M.~A., {Diamond}, P.~H., {Sagdeev}, R.~Z., {Aharonian}, F.~A., \&
  {Moskalenko}, I.~V. 2013, \apj, 768, 73

\bibitem[{{Mikhailovsky}(1967)}]{MikhailovskyChEnsk67}
{Mikhailovsky}, A.~B. 1967, Soviet Journal of Experimental and Theoretical
  Physics, 25, 623

\bibitem[{{Nayfeh}(1981)}]{Nayfeh81}
{Nayfeh}, A.~H. 1981, {Introduction to perturbation techniques}

\bibitem[{{O'C.~Drury}(2013)}]{DruryAnisotr13}
{O'C.~Drury}, L. 2013, ArXiv e-prints

\bibitem[{{Pauls} {et~al.}(1993){Pauls}, {Burger}, \& {Bieber}}]{Pauls93}
{Pauls}, H.~L., {Burger}, R.~A., \& {Bieber}, J.~W. 1993, International Cosmic
  Ray Conference, 3, 183

\bibitem[{{Ptuskin} {et~al.}(2008){Ptuskin}, {Zirakashvili}, \&
  {Plesser}}]{PtuskinNLDIFF08}
{Ptuskin}, V.~S., {Zirakashvili}, V.~N., \& {Plesser}, A.~A. 2008, Advances in
  Space Research, 42, 486

\bibitem[{{Schwadron} \& {Gombosi}(1994)}]{SchwadronTelegraph94}
{Schwadron}, N.~A., \& {Gombosi}, T.~I. 1994, \jgr, 99, 19301

\bibitem[{Vedenov {et~al.}(1962)Vedenov, Velikhov, \& Sagdeev}]{VVSQL62}
Vedenov, A.~A., Velikhov, E.~P., \& Sagdeev, R.~Z. 1962, NUCLEAR FUSION, 465

\end{thebibliography}

\end{document}